\documentclass[11pt]{article}

\date{}

\usepackage{color,soul}

\usepackage{bm}  

\usepackage[colorlinks,linkcolor=blue,citecolor=blue]{hyperref}
\usepackage{amsmath,amssymb,dsfont}
\usepackage{epsfig}
\usepackage{epstopdf}
\usepackage{xcolor,graphicx,float}
\usepackage{subfigure}
\usepackage{amsfonts}
\usepackage{amsmath}
\usepackage{multirow}

\usepackage{natbib}
\usepackage{appendix}
\usepackage{ulem}
\usepackage{color}
\usepackage{tabu}
\usepackage{hyperref}
\usepackage[hyphenbreaks]{breakurl}
\usepackage{booktabs}
\usepackage{threeparttable}

\newcommand{\E}{\mathbb{E}}
\newcommand{\1}{\mathds{1}}

\newcommand{\bea}{\bed\begin{array}{rl}}
\newcommand{\eea}{\end{array}\eed}

\newcommand{\al}{\alpha}

\newcommand{\rr}{{\hbox{{\rm I}{\kern -0.22em}{\rm R}}}}

\newtheorem{alg}{\noindent Algorithm}[section]
\newtheorem{thm}{\noindent Theorem}[section]

\newtheorem{lem}{\noindent Lemma}[section]

\newtheorem{example}{\noindent Example}[section]

\newcommand{\proof}{\noindent {\bf Proof} \hspace{2mm}}
\newcommand{\eproof}{$\quad \Box$}

\newtheorem{proposition}{Proposition}
\newtheorem{lemma}{Lemma}[section]
\newtheorem{definition}{Definition}[section]
\newtheorem{assum}{Assumption}[section]

\newtheorem{remark}{Remark}[section]

\newcounter{RomanNumber}

\begin{document}
\title{Semimartingale and continuous-time Markov chain approximation for rough stochastic
local volatility models \thanks{The work
was supported by National Natural Science Foundation of
China (Grant No. 12071373) and the Fundamental Research Funds for the Central Universities China (JBK1805001). We would like to thank Antoine Jacquier and Philipp Schoenbauer for discussions leading to improvement in the paper.  }}

\author{Jingtang Ma\thanks{School of Economic Mathematics, Southwestern University
of
Finance and Economics, Chengdu, 611130, P.R. China (Email: mjt@swufe.edu.cn). },
 Wensheng Yang\thanks{School of Economic Mathematics,
Southwestern University of Finance and Economics, Chengdu,
611130, P.R. China (Email: yangws@swufe.edu.cn). }
~and
 Zhenyu Cui\thanks{Corresponding author. School of Business, Stevens Institute of
 Technology, Hoboken,  New Jersey 07030, United States. (Email:
 zcui6@stevens.edu).} }
\date{\today}

\maketitle

\begin{abstract}
Rough  volatility models have recently been empirically shown to provide a good fit to historical volatility time series and implied volatility smiles of SPX options. They are continuous-time stochastic volatility models, whose volatility process is driven by a fractional Brownian motion with Hurst parameter less than half. Due to the challenge that it is neither a semimartingale nor a Markov process, there is no unified method that not only applies to all rough volatility models, but also is computationally efficient.
   This paper proposes a semimartingale and continuous-time Markov chain (CTMC) approximation approach for the general class of rough stochastic local volatility (RSLV) models. In particular, we introduce the perturbed stochastic local volatility (PSLV) model as the semimartingale approximation for the RSLV model and establish its existence, uniqueness and Markovian representation.  We propose a fast CTMC algorithm and prove its weak convergence. Numerical experiments demonstrate the accuracy and high efficiency of the method in pricing European, barrier and American options. Comparing with existing literature, a significant reduction in the CPU time to arrive at the same level of accuracy is observed.

\end{abstract}

\vspace{1.0cm}


\smallskip


\section{Introduction}

Recently, a new class of stochastic volatility model, named the \textit{rough volatility} model, was proposed in \cite{gatheral},  and has since then generated significant amount of interests from both academia and industry. The key insight of this model is to assume that  the latent stochastic volatility process is driven by a fractional Brownian motion, in contrast to a standard Brownian motion (e.g. in traditional stochastic (local) volatility models). The trajectories of the volatility process are continuous but exhibit irregular path properties due to the fractional Brownian motion driver.  From empirical studies, \cite{gatheral} find that the log-volatility essentially behaves like a fractional Brownian motion with Hurst exponent $H$ of order $0.1$, at any reasonable time scale. Further empirical evidence has been documented in \cite{fu2019}, where the authors constructed a quasi-likelihood estimator applied to realized volatility time series, and confirmed that the Hurst parameter is much smaller than half, i.e. volatility is indeed rough.

 The rough volatility model has enjoyed huge success in reproducing many stylized facts of historical volatility time series and implied volatility smiles for SPX options. On one hand, rough volatility models provide remarkably accurate fit to the shape of implied volatility smiles, and in particular for at-the-money skew curves. They also reproduce stylized facts for realized volatilities \citep{euch2019,Livieri}. On the other hand, it is consistent with  economic micro-structural models and naturally emerges from economic agents' behaviors, as shown in \cite{euch2018}. It also has intrinsic connections with Hawkes processes, see \cite{jaisson2016,Dandapani}.


The existence of the fractional kernel forces the variance process to leave both the semimartingale and Markovian worlds, hence one important yet challenging problem in the rough volatility research is to find an efficient and accurate method to evaluate derivative prices, whereas closed-form formulae are in general not available.  
 On one hand, several analytical approximation methods have been introduced and studied in \cite{forde2017,guennoun,forde2021}. On the other hand, the Monte Carlo simulation of rough volatility model has been studied in \cite{bayer,forde}, etc.  Due to the memory in the volatility process, the Monte Carlo simulation of rough volatility process is very time consuming, and there is recent research on improving its efficiency, see \cite{mc,bayer2020sparse,ma2021}.
 In the special class of affine rough volatility models, one is able to price options through Fourier transform based methods by utilizing the characteristic function via solving fractional Ricatti systems. In particular, \citet{euch} generalized the classical Heston model to the rough Heston model and derive the characteristic function of the log asset price. See also \cite{jaber-markov,richard,jaber-affine} for extensions.  Note that some numerical challenges still remain for solving the fractional Riccatti system in an efficient way, and see \cite{callegaro} for recent developments along that direction.  In general, without the affine structure, the characteristic function for the rough volatility model is not available.

In general, the Monte Carlo method has low computational efficiency, and the method of finding the characteristic function is not suitable for all rough volatility models outside of the special class of affine rough volatility models. Inspecting the existing literature\footnote{There is a website dedicated to collecting the most up-to-date literature on rough volatility research: \url{https://sites.google.com/site/roughvol/home/risks-1} }, to the best of authors' knowledge, there is no method that is not only generally applicable to various rough volatility models but also has good accuracy and computational efficiency.
 This motivates us to search for a general method that is widely applicable to the broad class of rough stochastic local volatility (RSLV) models (see equation \eqref{model:RSLV}).
 Inspired by the recent success on using the continuous-time Markov chain (CTMC) method in derivatives pricing (see the survey \cite{cui2019} and references therein), this paper aims to extend the applicability of the CTMC method to the realm of rough volatility models for the first time.

Our  method comprises of two main steps. The first step is a novel semimartingale approximation to the RSLV model, and we obtain the ``perturbed stochastic local volatility (PSLV) model", which is new to the literature. This removes the singularity in the kernel function. The second step is the CTMC approximation to the PSLV model, and we manage to obtain an explicit formula involving matrix expressions. In the first step above, we provide a new semimartingale approximation to the general class of RSLV models, which is of independent theoretical interest. It is important to distinguish this semimartingale approximation from a recent series of literature on the Markovian approximation to rough volatility models, which
started from  \cite{jaber-multi}, \cite{alfonsi}, and see also \cite{harms} for the case of fractional Brownian motion. The main idea there is to represent the fractional process as an integral over a family of Ornstein-Uhlenbeck processes, and then apply numerical discretization (i.e. quadrature) to the integral. The final result is a $(n+2)$-dimensional stochastic differential equation system, where $n$ is the number of grid points of numerical integration. In contrast, our approach is based on the perturbation idea, which was first introduced in \citet{dung} for the case of fractional Brownian motions, and we extend it to the case of general RSLV models to arrive at the PSLV model. The final result is a stochastic differential equation system, and the stochastic volatility process in this system is a semimartingale and Markov process. See Remark \ref{dim}. \vspace{0.2cm}






The contributions of this paper are three-fold:
\begin{enumerate}
  \item This paper extends the traditional stochastic local volatility (SLV) model to the rough version, and names it the  rough stochastic local volatility (RSLV) model. Using the semimartingale approximation to the RSLV model, a new model  named the ``perturbed stochastic local volatility" (PSLV) model, and its stochastic differential form are obtained.  
  In addition, this paper discusses the existence, uniqueness, regularity, semimartingale property and the Markov property of the PSLV model, and also prove that it converges weakly to the original RSLV model.

  \item A novel CTMC approximation method is developed, and we express options prices under the RSLV model in explicit matrix formulae. To the best of authors' knowledge, this is the first CTMC algorithm designed for the  RSLV model. Theoretical convergence of this algorithm is established. In addition, a fast algorithm (Algorithm \ref{alg:fast}) for pricing European and barrier option is given. Compared with the traditional coupled two-dimensional CTMC method \citep{cui-and-kirkby2018}, the new CTMC algorithm is decoupled. There is a significant improvement in computer storage space and computing capacity (see Remark \ref{re:ctmc}).

  \item Numerical examples demonstrate the  accuracy and high efficiency of our method.
  In particular, the method can deliver European and barrier options prices up to 3 digits of accuracy in 0.18 seconds of CPU time. Note that the method is universally fast and accurate across all RSLV models, and it is applicable not only to path-independent options such as European call/put options, but also to path-dependent options such as barrier options and American options. There is a  significant reduction in the CPU time to arrive at the same level of accuracy, as compared to benchmark methods in the literature.
\end{enumerate}

The remainder of this paper is organized as follows. Section \ref{sec:model} presents the new PSLV model, studies its properties, proves its convergence to the RSLV model and provides its Markovian representation. Section \ref{sec:ctmc} gives the CTMC approximation and establishes its weak convergence. Section \ref{sec:option}  considers the European, barrier and American options pricing problems under the RSLV model and gives the explicit matrix expressions for their prices. Numerical experiments are reported in Section \ref{sec:numerical}. Finally, Section \ref{section-conclusion} concludes the paper.

\section{Semimartingale approximation of rough stochastic local volatility
models}\label{sec:model}
\subsection{Rough stochastic local volatility models}\label{sec:rslv}
We consider the asset price $\{S_t:t\in\mathbb T\}$ with $\mathbb T:=[0,T]$ which is defined on a filtered
probability space $(\Omega,\mathcal{F},\mathbb{F},\mathbb{P})$, where
$\mathbb{F}=\{\mathcal{F}_t\}_{t\ge0}$ denotes the standard filtration generated by
a two-dimensional $\mathbb{F}-$Brownian motion $(B,B^{\perp})$  and $W=\rho B+\sqrt{1-\rho^2}B^{\perp}$ with constant
correlation $\rho\in (-1,1)$.
We first recall the classical stochastic local volatility (SLV) model as follows
 \begin{equation*}
\label{model:SLV}
\hbox{SLV:}\quad\left\{
 \begin{array}{ll}
dS_t = \mu(S_t,V_t) dt +\varphi(V_t)\nu (S_t) dW_t,\\
dV_t =b(V_t)dt
+\sigma(V_t)dB_t,
\end{array}
\right.
\end{equation*}
where $\mu(\cdot,\cdot):\mathbb R \times \mathbb R\rightarrow \mathbb R$, $\varphi(\cdot):\mathbb R\rightarrow \mathbb R_+$, $\nu(\cdot):\mathbb R\rightarrow \mathbb R_+$, $b(\cdot):\mathbb R\rightarrow\mathbb R$, $\sigma(\cdot):\mathbb R\rightarrow \mathbb R_+$.  $W_t$ and $B_t$ are two Brownian motions with a constant correlation $\rho\in(-1,1)$, that is, $E[dW_tdB_t]=\rho dt$. The SLV model nests several representative  models in the literature as special cases,  such as Heston \citep{heston}, $4/2$ \citep{grasselli}, Stein-Stein \citep{stein}, $3/2$ \citep{lewis}, Hull-White \citep{hull},  $\alpha$-Hypergeometric \citep{da},  SABR\footnote{Stochastic Alpha Beta Rho} \citep{hagan},  Heston-SABR \citep{van}, Quadratic SLV \citep{lipton}, etc. 

Recasting the classical SLV model into its rough correspondent, we have the following definition of the rough stochastic local volatility (RSLV) model.

\begin{definition}[rough stochastic local volatility model]
Under the risk-neutral measure, assume that the underlying asset price $S_t$
follows a rough stochastic local volatility (RSLV) model characterized by the
following two-dimensional diffusion system:
\begin{equation}
\label{model:RSLV}
\hbox{RSLV:}\quad\left\{
 \begin{array}{ll}
dS_t = \mu(S_t,V_t) dt +\varphi(V_t)\nu (S_t) dW_t,\\
V_t =V_0+\int^t_0K(t,s)\left(b(V_s)ds
+\sigma(V_s)dB_s\right),
\end{array}
\right.
\end{equation}
where $K(t,s):=\frac{(t-s)^{H-\frac{1}{2}}}{\Gamma(H+1/2)}$ is the fractional kernel with the Hurst parameter $H\in(0,1/2)$.
\end{definition}
 In order to ensure the strong existence of continuous solutions to \eqref{model:RSLV}, the following regularity assumption is necessary.
\begin{assum}\label{assum:mu}
$\mu(\cdot,\cdot):\mathbb R \times \mathbb R\rightarrow \mathbb R$, $\nu(\cdot):\mathbb R\rightarrow \mathbb R_+$, $b(\cdot):\mathbb R\rightarrow\mathbb R$, $\varphi(\cdot):\mathbb R\rightarrow \mathbb R_+$, $\sigma(\cdot):\mathbb R\rightarrow \mathbb R_+$  are all Lipschitz continuous functions with linear growth.
\end{assum}

\begin{proposition}\label{lemma:regularity}
Under Assumption \ref{assum:mu}, the equation system \eqref{model:RSLV} admits a unique strong continuous solution. Moreover $V_t$ and $S_t$ satisfies
\begin{equation*}
\sup_{t\in\mathbb T}\E[|V_t|^p]<\infty,~~\sup_{t\in\mathbb T}\E[|S_t|^p]<\infty,~~p>0.
\end{equation*}
In addition, $V$ and $S$ admit H\"{o}lder continuous paths on $\mathbb T$ of any order strictly less than $H$.
\end{proposition}
\proof
We refer to \citet{jaber-affine} for the proofs.
\eproof

\begin{remark}
This paper focuses on  kernel function of the following form $K(t,s)=\frac{(t-s)^{H-\frac{1}{2}}}{\Gamma(H+1/2)}$. However, under certain assumptions, the method in this paper can be applied directly to a more general class of kernel functions. We refer to \citet{jaber-affine} for discussions on the regularity condition of the kernel function.
\end{remark}

It is well known that the fractional kernel forces the variance process to leave both the semimartingale and Markovian worlds, which makes numerical approximation procedures a difficult and challenging task in practice. \citet{jaber-multi} use a multi-factor model to perform a Markovian approximation to RSLV. We now show an alternative approach to approximate RSLV by a semimartingale model through a perturbation idea.

\subsection{Semimartingale approximation}\label{sec:semi-approx}

Inspired by \citet{dung}, who performs a semimartingale approximation to the  fractional Brownian motion, we approximate the fractional kernel $K(t,s)$ by a perturbed kernel $K(t-\varepsilon,s)$ with $0<\varepsilon<<1$. This leads to the following approximation process $V_t^{\varepsilon}$ of the variance process $V_t$:
\begin{equation*}
V_t^{\varepsilon}:=V^\varepsilon_0+\int^t_0K(t+\varepsilon,s)\left(b(V^\varepsilon_s)ds
+\sigma(V^\varepsilon_s)dB_s\right),~~V^\varepsilon_0=V_0,
\end{equation*}
where
\begin{equation*}
K(t+\varepsilon,s)=\frac{(t+\varepsilon-s)^{H-\frac{1}{2}}}{\Gamma(H+1/2)}.
\end{equation*}

\begin{remark}
The original kernel function $K(t,s)$ is singular at the point $s=t$, and the kernel function $K(t+\varepsilon,s)$ obtained by thesemimartingale approximation is smooth for any $s\in[0,t]$.
\end{remark}

The first lemma establishes the semimartingale property of the process $V_t^{\varepsilon}$.
\begin{lemma}\label{lemma:semi}
For any $\varepsilon>0$, $V_t^{\varepsilon}$ is a $\mathcal{F}_t-$semimartingale with decomposition
\begin{equation}\label{eq:v-eps-phi}
V_t^{\varepsilon}=V^\varepsilon_0+\int^t_0\left(K(t+\varepsilon,s)b(V^\varepsilon_s)
+\psi^\varepsilon_s\right)ds+\int^t_0K(s+\varepsilon,s)\sigma(V^\varepsilon_s)dB_s,
\end{equation}
where
\begin{equation*}
\psi^\varepsilon_s:=\int^s_0\partial_1 K(s+\varepsilon,u)\sigma(V^\varepsilon_u)dB_u,~~~\partial_1 K(s+\varepsilon,u)=\frac{(H-\frac{1}{2})(s+\varepsilon-u)^{H-\frac{3}{2}}}{\Gamma(H+\frac{1}{2})}.
\end{equation*}
\end{lemma}
\proof
The lemma follows from a straightforward application of the stochastic Fubini
theorem (c.f. Theorem 2.2 in \citet{veraar}):
\begin{equation*}
\begin{aligned}
\int^t_0\psi^\varepsilon_sds
&=\int^t_0\int^s_0\partial_1 K(s+\varepsilon,u)\sigma(V^\varepsilon_u)dB_uds\\
&=\int^t_0\left(\int^t_u\partial_1 K(s+\varepsilon,u)ds\right)\sigma(V^\varepsilon_u)dB_u\\
&=\int^t_0\big(K(t+\varepsilon,u)-K(u+\varepsilon,u)\big)\sigma(V^\varepsilon_u)dB_u\\
&=\int^t_0K(t+\varepsilon,s)\sigma(V^\varepsilon_s)dB_s
-\int^t_0K(s+\varepsilon,s)\sigma(V^\varepsilon_s)dB_s.
\end{aligned}
\end{equation*}
This completes the proof.
\eproof

 \vspace{0.3cm}

The second lemma establishes the strong existence and uniqueness of $V_t^\varepsilon$.
\begin{lemma}\label{lemma:ex_uniq_semi}
Under Assumption \ref{assum:mu}, for any $\varepsilon>0$, there exists a unique strong solution $V_t^\varepsilon$. Moreover $V^\varepsilon_t$ satisfies
\begin{equation*}
\sup_{t\in\mathbb T}\E[|V^{\varepsilon}_t|^p]<\infty,~~p>0,
\end{equation*}
and admits H\"{o}lder continuous paths on $\mathbb T$ of any order strictly less than $H$.
\end{lemma}
\proof
We first prove the existence of $V^\varepsilon_t$. Thanks to the continuity of $K(t+\varepsilon,s)$ and boundedness of $\partial_1 K(t+\varepsilon,s)=\frac{(H-1/2)(t+\varepsilon-s)^{H-3/2}}{\Gamma(H+1/2)}$, the proof of Theorems 3.3 and 3.4 in \citet{jaber-affine} can be directly applied to prove the existence of $V^\varepsilon_t$.
Now we show the pathwise uniqueness. Since $V_t^\varepsilon$ is a semimartingale  as shown in Lemma \ref{lemma:semi} and $K(t+\varepsilon,t)=\frac{\varepsilon^{H-\frac{1}{2}}}{\Gamma(H+1/2)}<\infty$, we can use a similar proof as that of Proposition B.3. in \citet{jaber-multi} to establish the  uniqueness of $V_t^\varepsilon$. This completes the proof.
\eproof

 \vspace{0.3cm}

The next theorem proves the convergence of the semimartingale approximation.
\begin{thm}\label{thm:conver_v}
Under Assumption \ref{assum:mu}, the process $V^\varepsilon_t$ converges to $V_t$ in $\mathbf{L}^2(\Omega,\mathbb T)$ as $\varepsilon$ tends to $0$, uniformly in $t\in\mathbb T$.
\end{thm}
\proof
\begin{equation*}
\begin{aligned}
&E[|V^\varepsilon_t-V_t|^2]\\
=&E\Big[\Big|\int^t_0K(t+\varepsilon,s)\big(b(V^\varepsilon_s)-b(V_s)\big)ds
+\int^t_0\big (K(t+\varepsilon,s)-K(t,s)\big)b(V_s)ds\\
+&\int^t_0K(t+\varepsilon,s)\big)\big(\sigma(V^\varepsilon_s)-\sigma(V_s)\big)dB_s
+\int^t_0\big (K(t+\varepsilon,s)-K(t,s)\big)\sigma(V_s)dB_s\Big|^2\Big].
\end{aligned}
\end{equation*}
Recalling the power mean inequality: for $k\ge1,x_1,x_2,\dots,x_\ell\ge0,\ell\in\mathbb N_+$, $\sum^l_{i=1}\frac{x_i}{\ell}\le(\sum^l_{i=1}\frac{x^k_i}{\ell})^{1/k}$ and using It\^{o} isometry and the Cauchy-Schwarz's inequality, we have
\begin{equation*}
\begin{aligned}
&E[|V^\varepsilon_t-V_t|^2]\\
\le&4E\Big[\Big(\int^t_0K(t+\varepsilon,s)\big(b(V^\varepsilon_s)-b(V_s)\big)ds\Big)^2
+\Big(\int^t_0\big (K(t+\varepsilon,s)-K(t,s)\big)b(V_s)ds\Big)^2\\
&+\int^t_0K^2(t+\varepsilon,s)\big)\big(\sigma(V^\varepsilon_s)-\sigma(V_s)\big)^2ds
+\int^t_0\big (K(t+\varepsilon,s)-K(t,s)\big)^2\sigma^2(V_s)ds\Big]\\
\le&4E\Big[\int^t_0K^2(t+\varepsilon,s)\Big(
t\big(b(V^\varepsilon_s)-b(V_s)\big)^2+\big(\sigma(V^\varepsilon_s)-\sigma(V_s)\big)^2\Big)ds\\
&+\int^t_0 (K(t+\varepsilon,s)-K(t,s)\big)^2\big(tb^2(V_s)+\sigma^2(V_s)\big)ds\Big].
\end{aligned}
\end{equation*}
By the conditions that  $b$, $\sigma$ are Lipschitz continuous with linear growth, and the stochastic Fubini theorem, we have
\begin{equation*}
\begin{aligned}
E[|V^\varepsilon_t-V_t|^2]&\le
4\Big(C_1(1+t)\int^t_0K^2(t+\varepsilon,s)E[|V^\varepsilon_s-V_s|^2]ds\\
&+\int^t_0 (K(t+\varepsilon,s)-K(t,s)\big)^2C_1(1+t)\big(1+E[|V_s|]+E[V_s^2]\big)ds\Big).
\end{aligned}
\end{equation*}
Here and throughout this paper, we use $C_i$, $i=1,2,\dots$, to represent positive constants. By Taylor expansion, there is
\[
 K(t+\varepsilon,s)-K(t,s)=\frac{(t+\varepsilon-s)^{H-\frac{1}{2}}-(t-s)^{H-\frac{1}{2}}}{\Gamma(H+1/2)}
 =\varepsilon\frac{(H-\frac{1}{2})(t+\varepsilon-s)^{H-\frac{3}{2}}}{\Gamma(H+1/2)}+o(\varepsilon),\]
and $E[|V_s|^p]<\infty$ for $p>0$ from Proposition \ref{lemma:regularity}, then we obtain
\begin{equation*}
\begin{aligned}
&\int^t_0 \Big[(K(t+\varepsilon,s)-K(t,s)\big)^2C(1+t)\big(1+E[|V_s|]+E[V_s^2]\big)\Big]ds\\
&\le
C_2\int^t_0\Big( \varepsilon\frac{(H-\frac{1}{2})(t+\varepsilon-s)^{H-\frac{3}{2}}}{\Gamma(H+1/2)}+o(\varepsilon)\Big)ds
=C_2\varepsilon^{H+\frac{1}{2}}+\mathcal{O}(\varepsilon).
\end{aligned}
\end{equation*}
Finally, using Gr\"{o}nwall's inequality leads to
\begin{equation*}
\begin{aligned}
E[|V^\varepsilon_t-V_t|^2]&\le4\Big[C_1(1+t)\int^t_0K^2(t+\varepsilon,s)E[|V^\varepsilon_s-V_s|^2]ds+
C_2\varepsilon^{H+\frac{1}{2}}+\mathcal{O}(\varepsilon)\Big]\\
 &\le C_3\varepsilon^{H+\frac{1}{2}}.
\end{aligned}
\end{equation*}
This completes the proof.
\eproof

 \vspace{0.3cm}

Given the existence, uniqueness, semimartingale and convergence properties of $V_t^\varepsilon$, we now define the so-called perturbed stochastic local volatility (PSLV) model $(S^\varepsilon_t,V^\varepsilon_t)$, which serves as an approximation of $(S_t,V_t)$.

\begin{definition}[perturbed stochastic local volatility models]
We define the following stochastic local volatility model $(S^\varepsilon_t,V^\varepsilon_t)$ with perturbation parameters $0<\varepsilon<<1$  as the unique strong solution of
\begin{equation}
\label{model:PSLV}
\hbox{PSLV:}\quad\left\{
 \begin{array}{ll}
dS^\varepsilon_t = \mu(S^\varepsilon_t,V^\varepsilon_t) dt +\varphi(V^\varepsilon_t)\nu (S^\varepsilon_t) dW_t,\\
V_t^{\varepsilon}=V^\varepsilon_0+\int^t_0K(t+\varepsilon,s)\left(b(V^\varepsilon_s)ds
+\sigma(V^\varepsilon_s)dB_s\right),~~~V_0^{\varepsilon}=V_0,
\end{array}
\right.
\end{equation}
under the same filtered probability space $(\Omega,\mathcal{F},\mathbb{F},\mathbb{P})$ as $(S_t,V_t)$ defined by \eqref{model:RSLV}.
\end{definition}

\begin{proposition}\label{lemma:regularity_pslv}
Under Assumption \ref{assum:mu}, the equation system \eqref{model:PSLV} admits a unique strong continuous solution.
\end{proposition}
\proof
The strong existence and uniqueness of $V_t^\varepsilon$ is given by Lemma \ref{lemma:ex_uniq_semi}. Moreover, $S_t^\varepsilon$ satisfies a stochastic differential equation and there exists an unique strong solution to it under Assumption \ref{assum:mu} (see e.g., \citet{oksendal}). This completes the proof.
\eproof

\vspace{0.3cm}

 As shown in Theorem \ref{thm:conver_v}, $V_t^\varepsilon$ converges to $V_t$ as $\varepsilon$ tends to $0$. The next theorem shows the convergence of $S^\varepsilon_t$ to $S_t$.

\begin{thm}\label{thm:conver_s}
Under Assumption \ref{assum:mu}, the process $S^\varepsilon_t$ converges to $S_t$ in $\mathbf{L}^2(\Omega,\mathbb T)$ as $\varepsilon$ tends to $0$, uniformly in $t\in\mathbb T$.
\end{thm}
\proof
The proof of this theorem is similar to Theorem \ref{thm:conver_v}. Specifically, using power mean inequality, It\^{o} isometry and the Cauchy-Schwarz's inequality, we obtain
\begin{equation*}
\begin{aligned}
E[|S^\varepsilon_t-S_t|^2]&\le
2E\Big[\Big(\int^t_0\big(\mu(S^\varepsilon_s,V^\varepsilon_s)- \mu(S_s,V_s)\big)ds\Big)^2 +\int^t_0\big(\varphi(V^\varepsilon_s)\nu (S^\varepsilon_s)-\varphi(V_s)\nu (S_s)\big)^2 ds\Big]\\
&\le2E\Big[\int^t_0t\big(\mu(S^\varepsilon_s,V^\varepsilon_s)- \mu(S_s,V_s)\big)^2ds+\big(\varphi(V^\varepsilon_s)\nu (S^\varepsilon_s)-\varphi(V_s)\nu (S_s)\big)^2 ds\Big].
\end{aligned}
\end{equation*}
Then using the condition that $\mu$, $\varphi$ and $\nu$ are Lipschitz continuous with linear growth, and the stochastic Fubini theorem, we have
\begin{equation*}
\begin{aligned}
E[|S^\varepsilon_t-S_t|^2]&\le C_4(1+t)\Big(\int^t_0E[|S^\varepsilon_s-S_s|^2]ds+\int^t_0E[|V^\varepsilon_s-V_s|^2]ds\Big).
\end{aligned}
\end{equation*}
Thanks to Theorem \ref{thm:conver_v} and Gr\"{o}nwall's inequality, we have
\begin{equation*}
\begin{aligned}
E[|S^\varepsilon_t-S_t|^2]&\le C_5\varepsilon^{H+\frac{1}{2}}.
\end{aligned}
\end{equation*}
This completes the proof.
\eproof

\subsection{Markovian representation of the PSLV model}\label{sec:markovian}

In the PSLV model \eqref{model:PSLV} obtained by semimartingale approximation, since the singularity of the integral kernel $K$ at point $0$ is eliminated, we can prove that $V^{\varepsilon}_t$ is Markovian through the following theorem.
\begin{thm}\label{thm:dv}
The stochastic process $V^{\varepsilon}_t$ is a Markov process, and satisfies the following stochastic differential equation:
\begin{equation}\label{dif:v-ep-phi}
dV_t^{\varepsilon}=K^\varepsilon b(V^\varepsilon_t)dt+\varphi^\varepsilon_tdt
+K^\varepsilon \sigma(V^\varepsilon_t)dB_t,
\end{equation}
where
\[K^{\varepsilon}:=K(t+\varepsilon,t)\]
and
\begin{equation*}
\varphi^\varepsilon_t:=\int^t_0\partial_1 K(t+\varepsilon,u)\big(b(V^\varepsilon_t)dt+\sigma(V^\varepsilon_u)dB_u\big),~~~\partial_1 K(t+\varepsilon,u)=\frac{(H-\frac{1}{2})(t+\varepsilon-u)^{H-\frac{3}{2}}}{\Gamma(H+\frac{1}{2})}.
\end{equation*}
\end{thm}
\proof
Differentiating equation \eqref{eq:v-eps-phi}, \eqref{dif:v-ep-phi} is obtained directly. Now we show the Markovian property of $V^{\varepsilon}_t$. Integrating form $s$ to $t$, for $0\le s\le t\le T$,  both sides of the equation \eqref{dif:v-ep-phi} gives
\begin{equation*}
\begin{aligned}
&V_t^{\varepsilon}-V_s^{\varepsilon}\\
=&\int^t_sK(u+\varepsilon,u)\big(b(V^\varepsilon_u)du
+\sigma(V^\varepsilon_u)dB_u\big)+\int^t_s\int^u_0\partial_1 K(u+\varepsilon,\xi)\big(b(V^\varepsilon_\xi)d\xi+\sigma(V^\varepsilon_\xi)dB_\xi\big)du\\
=&\int^t_sK(u+\varepsilon,u)\big(b(V^\varepsilon_u)du
+\sigma(V^\varepsilon_u)dB_u\big)+\int^t_s\int^u_s\partial_1 K(u+\varepsilon,\xi)\big(b(V^\varepsilon_\xi)d\xi+\sigma(V^\varepsilon_\xi)dB_\xi\big)du\\
&+\int^t_s\int^s_0\partial_1 K(u+\varepsilon,\xi)\big(b(V^\varepsilon_\xi)d\xi+\sigma(V^\varepsilon_\xi)dB_\xi\big)du\\
=&\int^t_sK(u+\varepsilon,u)\big(b(V^\varepsilon_u)du
+\sigma(V^\varepsilon_u)dB_u\big)+\int^t_s\int^u_s\partial_1 K(u+\varepsilon,\xi)\big(b(V^\varepsilon_\xi)d\xi+\sigma(V^\varepsilon_\xi)dB_\xi\big)du\\
&+\int^t_s \big(K(s+\varepsilon,\xi)-K(\varepsilon,\xi)\big)\big(b(V^\varepsilon_\xi)d\xi+\sigma(V^\varepsilon_\xi)dB_\xi\big).
\end{aligned}
\end{equation*}
Thanks to the zero-mean property of the It\^{o} integral and Fubini theorem, there is
 \begin{equation*}
 \begin{aligned}
 E[V^{\varepsilon}_t|\mathcal{F}_s]=&
  E\bigg[V_s^{\varepsilon}+\int^t_s\bigg(K(u+\varepsilon,u)b(V^\varepsilon_u)+\int^u_s\partial_1 K(u+\varepsilon,\xi)b(V^\varepsilon_\xi)d\xi\\
  &+\big(K(s+\varepsilon,u)-K(\varepsilon,u)\big)b(V^\varepsilon_u)\bigg) du\bigg|\mathcal{F}_s\bigg].
 \end{aligned}
 \end{equation*}
According to Theorem 17.2.3 in \citet{cohen},  we have
 \begin{equation*}
 E[V^{\varepsilon}_t|\mathcal{F}_s]= E[V^{\varepsilon}_t|V^{\varepsilon}_s], ~~\text{for}~~0\le s\le t\le T,
 \end{equation*}
 and it follows that $V_t^{\varepsilon}$ is a Markov process for $t\in\mathbb{T}$. This completes the proof.
\eproof

\begin{remark}
It is well known that $V_t$ in the RSLV model \eqref{model:RSLV} is not a Markov process and does not have It\^{o} differential expression. The reason is that the integral kernel satisfies $K(t,t)=\infty$. However, after the semimartingale approximation, $V^{\varepsilon}_t$ in the PSLV model \eqref{model:PSLV}, which has a  smooth kernel and $K(t+\varepsilon,t)=\frac{\varepsilon^{H-\frac{1}{2}}}{\Gamma(H+1/2)}$ for a fixed $0<\varepsilon<<1$,  is a semimartingale, and it also has Ito differential expressions and Markov property.
\end{remark}

Although \eqref{dif:v-ep-phi} shows the It\^{o} differential form of $V^{\varepsilon}_t$, it is not conducive to calculation and simulation because the term $\psi^\varepsilon_t$ is still in the  It\^{o} integral form. Next, we consider another differential expression. Inspired by  \citet{jaber-multi}, the perturbed fractional kernel $K(t+\varepsilon,s)=\frac{(t+\varepsilon-s)^{H-\frac{1}{2}}}{\Gamma(H+1/2)}$ can be written as a Laplace transform of a positive measure $m$:
\begin{equation*}
K(t+\varepsilon,s)=\int^\infty_0e^{-\gamma(t+\varepsilon-s)}m(d\gamma),
~~~~m(d\gamma)=\frac{\gamma^{-H-\frac{1}{2}}d\gamma}{\Gamma(H+1/2)\Gamma(1/2-H)}.
\end{equation*}
Then by the stochastic Fubini theorem, we obtain that
\begin{equation}\label{def:v-vgamma}
V_t^{\varepsilon}=V^\varepsilon_0+\int^\infty_0e^{-\gamma\varepsilon}
V_t^{\varepsilon,\gamma}m(d\gamma),
\end{equation}
where
\begin{equation}\label{def-v-eps}
V_t^{\varepsilon,\gamma}:=\int^t_0e^{-\gamma(t-s)}\left(b(V^\varepsilon_s)ds
+\sigma(V^\varepsilon_s)dB_s\right).
\end{equation}

\begin{thm}\label{thm:dv}
The PSLV model \eqref{model:PSLV} can be expressed as the following system stochastic differential equation:
\begin{equation}\label{dif:v-ep}
\left\{
 \begin{array}{ll}
dS^\varepsilon_t = \mu(S^\varepsilon_t,V^\varepsilon_t) dt +\varphi(V^\varepsilon_t)\nu (S^\varepsilon_t) dW_t,\\
dV^\varepsilon_t =-\left(\int^\infty_0e^{-\gamma\varepsilon}
\gamma V_t^{\varepsilon,\gamma}m(d\gamma)\right)dt+K^\varepsilon b(V^\varepsilon_t)dt
+K^\varepsilon \sigma(V^\varepsilon_t)dB_t,\\
dV_t^{\varepsilon,\gamma}=\left(-\gamma V_t^{\varepsilon,\gamma}+b(V^\varepsilon_t)\right)dt
+\sigma(V^\varepsilon_t)dB_t,
\end{array}
\right.
\end{equation}
\end{thm}
 \proof
 By It\^{o}'s lemma, for fixed $\gamma$, we have
 \begin{equation*}
 \begin{aligned}
 dV_t^{\varepsilon,\gamma}&=d\left(\int^t_0e^{-\gamma
(t-s)}\left(b(V^{\varepsilon}_s)ds +\sigma(V^{\varepsilon}_s)dB_s\right)\right)\\
&=-\gamma \left(\int^t_0e^{-\gamma
(t-s)}\left(b(V^{\varepsilon}_s)ds +\sigma(V_s)dB_s\right)\right)dt+b(V^{\varepsilon}_t)dt +\sigma(V^{\varepsilon}_t)dB_t\\
&=\left(-\gamma V_t^{\varepsilon,\gamma} +b(V^{\varepsilon}_t)\right)dt+\sigma(V^{\varepsilon}_t)dB_t.
\end{aligned}
\end{equation*}
From the definition of $V_t^{\varepsilon,\gamma}$ \eqref{def-v-eps}, we have
\begin{equation*}
\begin{aligned}
E[|V_t^{\varepsilon,\gamma}|]
&=E\left[\left|\int^t_0e^{-\gamma(t-s)}\left(b(V^\varepsilon_s)ds
+\sigma(V^\varepsilon_s)dB_s\right)\right|\right]\\
&\le E\left[\int^t_0e^{-\gamma(t-s)}\big(\left|b(V^\varepsilon_s)\right|ds+\left|\sigma(V^\varepsilon_s)\right|dB_s\big)\right].
\end{aligned}
\end{equation*}
Since $b(\cdot):\mathbb R\rightarrow \mathbb R$ and $\sigma(\cdot):\mathbb R\rightarrow \mathbb R_+$ are all Lipschitz continuous functions with linear growth and $\sup\limits_{t\in\mathbb T}\E[|V^{\varepsilon}_t|^p]<\infty,~~p>0$, then by Fubini theorem, we have that
\begin{equation*}
\begin{aligned}
E[|V_t^{\varepsilon,\gamma}|]
&\le C \int^t_0e^{-\gamma(t-s)}ds=C\frac{1-e^{-\gamma t}}{\gamma}.
\end{aligned}
\end{equation*}
 We use the Lebesgue dominated convergence theorem to rewrite \eqref{def:v-vgamma} as
\begin{equation*}
\begin{aligned}
dV_t^{\varepsilon}&=\int^\infty_0e^{-\gamma\varepsilon}
dV_t^{\varepsilon,\gamma}m(d\gamma)\\
&=-\left(\int^\infty_0e^{-\gamma\varepsilon}
\gamma V_t^{\varepsilon,\gamma}m(d\gamma)\right)dt+\int^\infty_0e^{-\gamma\varepsilon}
m(d\gamma)b(V^\varepsilon_t)dt
+\int^\infty_0e^{-\gamma\varepsilon}
m(d\gamma)\sigma(V^\varepsilon_t)dB_t\\
&=-\left(\int^\infty_0e^{-\gamma\varepsilon}
\gamma V_t^{\varepsilon,\gamma}m(d\gamma)\right)dt+K^\varepsilon b(V^\varepsilon_t)dt
+K^\varepsilon \sigma(V^\varepsilon_t)dB_t.
\end{aligned}
\end{equation*}
Note that by comparing with the formula \eqref{dif:v-ep-phi}, it can be seen that the first term in the right of above formula is actually equal to $\varphi^\varepsilon_t$ in \eqref{dif:v-ep-phi} by using Laplace transform to $\partial_1 K(t+\varepsilon,u)$. This completes the proof.
 \eproof

\vspace{0.3cm}

\begin{remark}\label{dim}
Based on the multifactor approximation, similar It\^{o} differential expressions with Markov properties can be obtained under the rough Heston model (see formula (1.4) in \citet{jaber-multi}). It is worth noting that their approximation method obtains an $(n+2)$-dimensional model, where $n$ is the number of grid points of numerical integration and the multifactor model converges to the original model as $n$ tends to infinity. In contrast, the approximate process $V_t^{\varepsilon}$ with Markov property can be obtained by the semimartingale approximation without numerical integration, thereby avoiding the difficulties caused by multi-dimensional problems in simulation and calculation.
\end{remark}

Next we consider how to use the Markov chain approximation methods to solve this stochastic differential equation system.

\section{CTMC approximation}\label{sec:ctmc}
In this section, we use the CTMC method introduced in \citet{mija} to approximate $(S^\varepsilon_t,V^\varepsilon_t)$ defined in  \eqref{model:PSLV} by a continuous-time Markov chain. To simplify the analysis, we first decouple the correlation between the two driving Brownian motions by introducing an auxiliary process $X^\varepsilon_t$ in the following lemma.
\begin{lemma}\label{lem:AuxLem}
Define the functions $g(x):=\int_{\cdot}^x \frac{1}{\nu(u)}du$ and $f(x):=\int_{\cdot}^x \frac{\varphi(u)}{K^\varepsilon \sigma(u)}du$. Then the dynamics in \eqref{model:PSLV} can be rewritten as
\begin{equation}\label{model:Aux}
\left\{
 \begin{array}{ll}
dX^\varepsilon_t=\theta(X^\varepsilon_t,V^\varepsilon_t) dt+ \sqrt{1-\rho^2}\varphi(V^\varepsilon_t)dB^{\bot}_t,\\
dV^\varepsilon_t =-\left(\int^\infty_0e^{-\gamma\varepsilon}
\gamma V_t^{\varepsilon,\gamma}m(d\gamma)\right)dt+K^\varepsilon b(V^\varepsilon_t)dt
+K^\varepsilon \sigma(V^\varepsilon_t)dB_t,\\
dV_t^{\varepsilon,\gamma}=\left(-\gamma V_t^{\varepsilon,\gamma}+b(V^\varepsilon_t)\right)dt
+\sigma(V^\varepsilon_t)dB_t,
\end{array}
\right.
\end{equation}
where the dynamics of the auxiliary process $X^\varepsilon_t:=g(S^\varepsilon_t) - \rho f(v^\varepsilon_t)$ and the standard Brownian motion $B^{\bot}_t:=\frac{W_t-\rho B_t}{\sqrt{1-\rho^2}}$  is independent from $B_t$ with a constant
correlation $\rho\in (-1,1)$. Here
\begin{equation}
\begin{aligned}\label{def-theta}
\theta(X^\varepsilon_t,V^\varepsilon_t)&:=\frac{\mu(S^\varepsilon_t,V^\varepsilon_t)}{\nu(S^\varepsilon_t)}
-\frac{\nu^{\prime}(S^\varepsilon_t)}{2}\varphi^2(V^\varepsilon_t) - \frac{\rho}{2}K^\varepsilon \big(\sigma(V^\varepsilon_t)\varphi^{\prime}(V^\varepsilon_t) - \sigma'(V^\varepsilon_t)\varphi(V^\varepsilon_t)\big)\\
&-\rho \left(-\int^\infty_0e^{-\gamma\varepsilon}
\gamma V_t^{\varepsilon,\gamma}m(d\gamma)+K^\varepsilon b(V^\varepsilon_t)\right)
\frac{\varphi(V^\varepsilon_t)}{K^\varepsilon \sigma(V^\varepsilon_t)}.
\end{aligned}
\end{equation}
\end{lemma}
\proof
This proof follows similarly from  Lemma 1 in  \citet{cui-and-kirkby2018}.
\eproof

\subsection{The construction of the CTMC approximation}
We first recall the basic setup of CTMC. Denote
\begin{equation*}
\begin{aligned}
\mathcal{M}&:=\{1,2,\ldots,M\},~~~~\mathcal{M}^\circ:=\{2,\ldots,M-1\},\\
\mathcal{N}&:=\{1,2,\ldots,N\},~~~~~\mathcal{N}^\circ:=\{2,\ldots,N-1\}.
\end{aligned}
\end{equation*}
Recall that a stochastic process $\al(t)$ taking values in the set
$\mathcal{M}$ of M possible states is a CTMC if the distribution of
$\al(t + \Delta t)$, conditioned on the current state and the past history up to
time $t$, depends only on the current state $\alpha(t)$. The transition dynamics of
$\alpha(t)$ are characterized by the rate matrix ${\bm Q}= (q_{ij})_{M\times
M}$, whose elements $q_{ij}$ satisfy (i) $q_{ii}\leq 0$, and
$q_{ij}\geq 0$, if $i\neq j$, and
(ii) $\sum_{j} q_{ij} = 0$, $\forall i\in\mathcal{M}$.

In terms of
$q_{ij}$, for a time increment $\Delta t\geq 0$, the transition probability
matrix ${\bf P}(\Delta t)$ has the following matrix exponential representation:
\[
 {\bf P}(\Delta t) = \exp({\bm Q}\cdot\Delta t):=\sum_{i=0}^{\infty}
 \frac{({\bm Q}\cdot\Delta t)^k}{k!},
\]
 with elements $p_{ij}(\Delta t):=\hbox{Prob}\left\{\alpha(t + \Delta t)=j \mid\alpha(t)=i\right\}$.

We first derive the CTMC approximation $\widetilde V^{\varepsilon}_t\in\{v^{\varepsilon}_1,v^{\varepsilon}_2,
\dots,v^{\varepsilon}_{M}\}$ of the variance process $V^{\varepsilon}_{t}$ over a general non-uniform grid $\{v^{\varepsilon}_i\}_{i=1}^{M}$ for $i\in\mathcal{M}$. Recall \eqref{def:v-vgamma} and let
\begin{equation*}
\widetilde V_t^{\varepsilon}=\widetilde V^\varepsilon_0+\int^\infty_0e^{-\gamma\varepsilon}
\widetilde V_t^{\varepsilon,\gamma}m(d\gamma),
\end{equation*}
where $\widetilde V^\varepsilon_0:=V^\varepsilon_0$, $\widetilde V_t^{\varepsilon,\gamma}\in \{v^{\varepsilon,\gamma}_i\}_{i=1}^{M}$
with $v^{\varepsilon,\gamma}_i=\frac{v^{\varepsilon}_i-V^\varepsilon_0}{\int^\infty_0e^{-\gamma\varepsilon} e^{-\gamma}m(d\gamma)}e^{-\gamma}=:\frac{v^{\varepsilon}_i-V^\varepsilon_0}{R}e^{-\gamma}$ and
\begin{equation*}
\hbox{Prob}\left\{\widetilde V_t^{\varepsilon,\gamma}=v^{\varepsilon,\gamma}_i \big| \widetilde V_t^{\varepsilon}=v^{\varepsilon}_i\right\}=1.
\end{equation*}

\begin{remark}
The relationship between $V_t^{\varepsilon,\gamma}$ and $V_t^{\varepsilon}$ is given by the equation \eqref{def:v-vgamma}. Note that the solution of the integral equation \eqref{def:v-vgamma} exists and is unique (see \citet{jaber-multi}). Thus there is a one-to-one correspondence between $V_t^{\varepsilon}$ and $V_t^{\varepsilon, \gamma}$.  Recall that $V_t^{\varepsilon}$ is Markov, hence we can use a CTMC to approximate it, and the corresponding finite state space is ${v_i^{\varepsilon}}$. Note that the intermediate auxiliary process $V_t^{\varepsilon, \gamma}$ is not Markov, and we are not constructing a CTMC approximation to it.  Hence $v^{\varepsilon,\gamma}_i$ should not be interpreted as the grid corresponding to a CTMC, but it is rather solved from the integral equation \eqref{def:v-vgamma} when a value $v_i^{\varepsilon}$ is substituted into that equation.  Moreover, the variable $\gamma$ in $V_t^{\varepsilon, \gamma}$ (defined by \eqref{def-v-eps}) appears in the exponential form, which is the reason why $v^{\varepsilon,\gamma}_i$ is set to the form $\frac{v^{\varepsilon}_i-V^\varepsilon_0}{R}e^{-\gamma}$. As a by-product, for a given grid of the CTMC approximating  $V_t^{\varepsilon}$, we have a uniquely defined corresponding value for $V_t^{\varepsilon, \gamma}$.  Knowing this one-to-one link between a realization of the Markov process $V_t^{\varepsilon}$ and the non-Markovian process $V_t^{\varepsilon, \gamma}$ is important, and is crucial for the design of the CTMC approximation to $V_t^{\varepsilon}$. Recall from \eqref{dif:v-ep} that the drift term of $V_t^{\varepsilon}$ contains an integral with respect to $V_t^{\varepsilon, \gamma}$. Based on the above link, when we carry out moment matching to construct the CTMC approximation to $V_t^{\varepsilon}$, we can actually express the drift term as an explicit function of $v_i^{\varepsilon}$ and separate the integral with respect to $\gamma$ into a  separate constant term $R$. This is the key advantage of the CTMC method as we avoid the discretization of the integral with respect to $\gamma$ through a quadrature method, and this fact precisely leads to a dimension reduction. Essentially, the process $V_t^{\varepsilon, \gamma}$  is just an intermediate auxiliary process that is uniquely characterized through the integral equation  \eqref{def:v-vgamma}. It does not need to be Markov, and the property of this intermediate process does not affect our construction of the CTMC approximation. We construct the CTMC approximation only to the $X_t^{\varepsilon}$ and $V_{t}^{\varepsilon}$, but not $V_t^{\varepsilon, \gamma}.$ 
To sum up, we use a CTMC $\widetilde V^{\varepsilon}_t$ to approximate $V_t^{\varepsilon}$ based on its Markov property. We first establish the grid points $v_i^{\varepsilon}$, and then use the integral equation \eqref{def:v-vgamma} to solve $v_i^{\varepsilon,\gamma}$ in terms of $v_i^{\varepsilon}$, and finally substitute it into the drift term of the SDE \eqref{dif:v-ep} of $V_t^{\varepsilon}$ and set up moment matching equations  to obtain the generator matrix of $\widetilde V^{\varepsilon}_t$.
\end{remark}

According to \citet{mija} and the SDE of $V_t^\varepsilon$ in \eqref{dif:v-ep}, the elements of the tridiagonal generator matrix $\mathbf Q=(q_{ij})_{M\times M}$ of $\widetilde V^{\varepsilon}_t$ for $i\in\mathcal{M}^\circ$, $j\in\mathcal{M}$, are uniquely determined through the following system of equations:
\begin{equation}\label{q-MCA-v}
\left[
\begin{array}{ccc}
1& 1& 1 \\
-h_{i}& 0& h_{i+1}\\
 h^2_{i}& 0& h^2_{i+1}
\end{array}
\right]
\left[
\begin{array}{c}
q_{i,i-1} \\
q_{i,i}\\
q_{i,i+1}
\end{array}
\right]
=
\left[
\begin{array}{c}
0 \\
(v^{\varepsilon}_i-V^\varepsilon_0)\widehat{R}+K^\varepsilon b(v^{\varepsilon}_i)\\
(K^\varepsilon)^2\sigma^2(v^{\varepsilon}_i)
\end{array}
\right],
\end{equation}
where $h_i=v^{\varepsilon}_i-v^{\varepsilon}_{i-1}$, $\widehat{R}:=-\int^\infty_0e^{-\gamma\varepsilon}
\gamma e^{-\gamma}m(d\gamma)/R$, $q_{1,j}=q_{M,j}=0$, $q_{i,j} = 0$ for
$|i-j|>1$, $i,j\in\mathcal{M}$. Solving \eqref{q-MCA-v} gives
\begin{equation}\label{q-solve}
\left\{
\begin{array}{l}
q_{i,i-1}=\frac{(K^\varepsilon)^2  \sigma^2(v^{\varepsilon}_i)
-((v^{\varepsilon}_i-V^\varepsilon_0)\widehat{R}+K^\varepsilon b(v^{\varepsilon}_i))h_{i+1}}{h_{i}(h_{i+1}+h_{i})},\\
q_{i,i}=\frac{-(K^\varepsilon)^2  \sigma^2(v^{\varepsilon}_i)
+((v^{\varepsilon}_i-V^\varepsilon_0)\widehat{R}+K^\varepsilon b(v^{\varepsilon}_i))(h_{i+1}-h_{i})}{h_{i+1}h_{i}},\\
q_{i,i+1}=\frac{(K^\varepsilon)^2  \sigma^2(v^{\varepsilon}_i)
+((v^{\varepsilon}_i-V^\varepsilon_0)\widehat{R}+K^\varepsilon b(v^{\varepsilon}_i))h_{i}}{h_{i+1}(h_{i+1}+h_{i})}.
\end{array}
\right.
\end{equation}
Thus the Markov process $V^{\varepsilon}_{t}$ from \eqref{model:Aux} is approximated by a continuous-time Markov chain $\widetilde{V}^{\varepsilon}_{t}$ with the generator matrix ${\bm Q}=(q_{i,j})_{M\times M}$, whose entries are given in \eqref{q-solve}.
\begin{remark}
It is worth noting that the generator matrix ${\bm Q}=(q_{i,j})_{M\times M}$ of $\widetilde{V}^{\varepsilon}_{t}$ is well-defined and time-homogeneous.  Each element $q_{i,j}$ shown in \eqref{q-solve} is explicitly expressed, where $v^{\varepsilon}_i$ and $h_{i}$ are provided by grids design, $R$ and $\widehat{R}$  are both constants, and $K^\varepsilon =\varepsilon^{H-\frac{1}{2}}/\Gamma(H+1/2)$. Therefore, the CTMC method is fully explicit and very computational friendly.
\end{remark}

Next, we derive the CTMC approximation $\widetilde X^{\varepsilon}_t\in\{x^{\varepsilon}_1,x^{\varepsilon}_2,
\dots,x^{\varepsilon}_{N}\}$ of the auxiliary process $X^{\varepsilon}_{t}$ over a general non-uniform grid $\{x^{\varepsilon}_i\}_{i=1}^{N}$, where $x^{\varepsilon}_i=\sum^i_{\ell=1}\delta_\ell$ for $i\in\mathcal{N}$, and $\delta_\ell$ is the grid interval. After approximating the variance process $V^{\varepsilon}_t$ by the CTMC $\widetilde V^{\varepsilon}_t$, the auxiliary process becomes a nonlinear regime-switching diffusion and its parameters have $M$ states. In this way, we can use the technique introduced in \citet{cui-and-kirkby2018} to approximate $X^{\varepsilon}_t$ by a continuous-time Markov chain $\widetilde X^{\varepsilon}_t$. In particular, according to \eqref{model:Aux}, for each $\ell \in \mathcal{M}$, we define a second-layer Markov chain approximation which is determined by the rate  matrix ${\bf \Lambda}_\ell=(\lambda^\ell_{ij})_{N\times N}$ with

\begin{equation}\label{q-MCA-LoSkin1}
\left[
\begin{array}{ccc}
1& 1& 1 \\
-\delta_{i}& 0& \delta_{i+1}\\
 \delta_{i}^2& 0& \delta_{i+1}^2
\end{array}
\right]
\left[
\begin{array}{c}
\lambda^\ell_{i,i-1} \\
\lambda^\ell_{i,i}\\
\lambda^\ell_{i,i+1}
\end{array}
\right]
=
\left[
\begin{array}{c}
0 \\
\theta(x^\varepsilon_i, v^\varepsilon_l)\\
(1-\rho^2)\varphi^2(v^\varepsilon_l)
\end{array}
\right],
\end{equation}
where  $\lambda^\ell_{1i}=\lambda^\ell_{Ni}=0$
and $\lambda^\ell_{i,j} = 0$ for $|i-j|>1$,  $i,j\in\mathcal{N}$ and $\ell\in\mathcal{M}$. Solving \eqref{q-MCA-LoSkin1} gives
\begin{equation}\label{def-lambda}
\left\{
\begin{array}{l}
\lambda^\ell_{i,i-1}=\frac{(1-\rho^2)\varphi^2(v^\varepsilon_\ell)-\theta(x^\varepsilon_i, v^\varepsilon_l)\delta_{i+1}}{\delta_{i}(\delta_{i+1}+\delta_{i})}, \\ \lambda^\ell_{i,i}=\frac{-(1-\rho^2)\varphi^2(v^\varepsilon_\ell)+\theta(x^\varepsilon_i, v^\varepsilon_l)(\delta_{i+1}-\delta_{i})}{\delta_{i+1}\delta_{i}},\\
\lambda^\ell_{i,i+1}=\frac{(1-\rho^2)\varphi^2(v^\varepsilon_\ell)+\theta(x^\varepsilon_i, v^\varepsilon_l)\delta_{i}}{\delta_{i+1}(\delta_{i+1}+\delta_{i})}.
\end{array}
\right.
\end{equation}
Note that $\lambda^\ell_{i,j}$ is well-defined for  $i,j\in\mathcal{N}$, $\ell\in\mathcal{M}$, where
\begin{equation*}
\begin{aligned}
\theta(x^\varepsilon_i, v^\varepsilon_\ell)&=\frac{\mu(s^\varepsilon_{i,\ell},v^\varepsilon_\ell)}{\nu(s^\varepsilon_{i,\ell})}
-\frac{\nu^{\prime}(s^\varepsilon_{i,\ell})}{2}\varphi^2(v^\varepsilon_\ell) + \frac{\rho}{2}\big(\sigma(v^\varepsilon_\ell)\varphi^{\prime}(v^\varepsilon_\ell) - \sigma'(v^\varepsilon_\ell)\varphi(v^\varepsilon_\ell)\big)\\
&-\rho\left((v^\varepsilon_\ell-V^\varepsilon_0)\widehat{R}+\frac{\varepsilon^{H-\frac{1}{2}}b(v^\varepsilon_\ell)}{\Gamma(H+1/2)}\right)
\frac{\varphi(v^\varepsilon_\ell)\Gamma(H+1/2)}{\varepsilon^{H-\frac{1}{2}}\sigma(v^\varepsilon_\ell)},
\end{aligned}
\end{equation*}
with $s^\varepsilon_{i,\ell}=g^{-1}(x^\varepsilon_i+\rho f(v^\varepsilon_\ell))$.

According to \citet{song2016}, $(\widetilde X^{\varepsilon}_t,\widetilde V^{\varepsilon}_{t})$ can be represented as a one-dimensional CTMC with a $NM \times NM$  transition rate matrix:
\begin{equation}
\begin{aligned}
{\bf \Lambda}&=\left( \begin{array}{cccc}
q_{11}{\bf I}_N+ {\bf \Lambda}_1 & q_{12}{\bf I}_N & \cdots & q_{1 M}{\bf I}_N \\
q_{21}{\bf I}_N & q_{22}{\bf I}_N+ {\bf \Lambda}_2 & \cdots &  q_{2 M}{\bf I}_N\\
\vdots & \vdots & \ddots & \vdots \\
q_{M 1}{\bf I}_N & q_{M 2}{\bf I}_N & \cdots & q_{M M}{\bf I}_N+ {\bf \Lambda}_{M} \end{array} \right),\label{matrix}
\end{aligned}
\end{equation}
where ${\bf I}_N$ is the $N\times N$ identity matrix, $q_{i,j}$ and ${\bf \Lambda}_l=(\lambda^l_{i,j})_{N\times N}$ are respectively given by \eqref{q-solve} and \eqref{def-lambda}. Recall \eqref{model:Aux} and denote
\[\widetilde S^\varepsilon_t:=g^{-1}\big(\widetilde X^\varepsilon_t+\rho f(\widetilde V^\varepsilon_t)\big).\]
Then for any continuous function $\phi$, with the setting that $g(s^\varepsilon)-\rho f(v^\varepsilon)=x^\varepsilon=x^\varepsilon_i$, $v^\varepsilon=v^\varepsilon_l$,
\begin{equation}\label{ctmc-formula}
\begin{aligned}
&\E\left[ \phi(S^\varepsilon_T,V^\varepsilon_T)|S^\varepsilon_t=s^\varepsilon,V^\varepsilon_t=v^\varepsilon\right]\\
=&\E\left[ \phi(g^{-1}\big(X^\varepsilon_t+\rho f( V^\varepsilon_t)\big),V^\varepsilon_T)|X^\varepsilon_t=x^\varepsilon,V^\varepsilon_t=v^\varepsilon\right]\\
\approx& \E\left[ \phi(g^{-1}\big(\widetilde X^\varepsilon_t+\rho f(\widetilde V^\varepsilon_t)\big),\widetilde V^\varepsilon_T)|\widetilde X^\varepsilon_t=x^\varepsilon_i,\widetilde V^\varepsilon_t=v^\varepsilon_l\right]\\
=&\mathbf{e}_{i,l} \cdot \exp (\mathbf{\Lambda}(T-t)) \cdot \mathbf{\Phi},
\end{aligned}
\end{equation}
where ${\bf e}_{i,l}$ is a $1\times MN$ vector with all entries equal to $0$ except that the $(l-1)N+i$ entry is equal to $1$, ${\bf \Lambda}$ is given by \eqref{matrix}, and the payoff vector ${\bf \Phi}$ an $MN\times 1$ vectors with elements ${ \bf \Phi}_{(l-1)N+i}= \phi(g^{-1}\big(x^\varepsilon_i+\rho f(v^\varepsilon_l)\big),v^\varepsilon_l)$, for $l\in\mathcal{M}$, $i\in\mathcal{N}$.

\subsection{Convergence analysis for the CTMC approximation}

Denote $\mathbb S:=[s_{\min},s_{\max}]$ for $-\infty< s_{\min}<s_{\max}<\infty$. Let $\mathbb S^\varepsilon:=[s^\varepsilon_{\min},s^\varepsilon_{\max}]$, $\mathbb V^\varepsilon:=[v^\varepsilon_{\min},v^\varepsilon_{\max}]$, $\mathbb X^\varepsilon:=[x^\varepsilon_{\min},x^\varepsilon_{\max}]$ and  $\mathbb T:=[0,T]$  be the range of the asset process $S^\varepsilon_t$, volatility process $V^\varepsilon_t$, auxiliary process $X^\varepsilon_t$ and time $t$, where $-\infty< s^\varepsilon_{\min}<s^\varepsilon_{\max}<\infty$, $-\infty< v^\varepsilon_{\min}<v^\varepsilon_{\max}<\infty$, $x^\varepsilon_{\min}=g(s^\varepsilon_{\min})+\rho f(v^\varepsilon_{\min})$ and $x^\varepsilon_{\max}=g(s^\varepsilon_{\max})+\rho f(v^\varepsilon_{\max})$. The convergence analysis of the CTMC approximation is based on the following lemma.

\begin{lem}\label{lem:ctmc-semi}
Let $S_t\in\mathbb{S}$ for $t\in\mathbb{T}$ be a Feller process whose infinitesimal generator is given by
 \[\mathcal{L}G(s):=\lim_{t\to0}\frac{\E[G(S_t)|S_0=s]-G(s)}{t}.\]
Let $\widetilde S_t^{n}\in {\mathbb S}^{n}$ be the continuous-time Markov chain with the generator ${\mathcal{L}}^n$, ${\mathbb S}^{n}\subseteq{\mathbb S}$ and when n tends to infinity, ${\mathbb S}^{n}={\mathbb S}$. Assume that for each $G\in \mathcal{C}^2({\mathbb S})$,
\begin{equation}\label{con-condition}
\lim_{n\to\infty}\max_{s\in{\mathbb S}^n}|\mathcal{L}G(s)-{\mathcal{L}}^nG(s)|=0.
\end{equation}
Then $\widetilde S_t^{n}$ converges weakly to $S_t$ as $n$ goes to infinity.
\end{lem}
\proof
According to Section 8.7 in \citet{durrett} and Theorem 10.1.1 in \citet{kushner}, it can be deduced from condition \eqref{con-condition} that $\widetilde S_t^{n}$ is tight. Then by Theorem 4.2.11 in \citet{ethier}, we obtain the weak convergence of $\widetilde S_t^{n}$ to $S_t$. This completes the proof.
\eproof

\vspace{0.3cm}

The first theorem gives the weak convergence of $\widetilde V^{\varepsilon}_t$  to $V^{\varepsilon}_t$.
\begin{thm}\label{thm:con-ctmc-vgamma}
Assuming that the grid interval $h$ satisfies: $h_\ell=\mathcal{O}(M^{-1})$, $|h_\ell-h_{\ell+1}|\le \mathcal{O}(M^{-2})$ and  Assumption \ref{assum:mu} holds, the continuous-time Markov chain $\widetilde{V}^{\varepsilon}_{t}$ with the generator ${\bm Q}=(q_{i,j})_{M\times M}$ converges weakly to ${V}^{\varepsilon}_{t}$, as $M$ goes to infinity.
\end{thm}
\proof
Theorem \ref{thm:dv} shows the Markov property of $V_t^{\varepsilon}$. According to the continuity of Riemann integral, $V_t^{\varepsilon}$ in \eqref{model:PSLV} has continuous simple path, therefore is a Feller process. Here we assume that $\hbox{Prob}\left\{V_t^{\varepsilon,\gamma}=v^{\varepsilon,\gamma}_\ell \big| V_t^{\varepsilon}=v^{\varepsilon}_\ell\right\}=1.$
Moreover, for $G\in \mathcal{C}^2({\mathbb V}^\varepsilon)$, $\ell\in\mathcal{M}^\circ$,
\begin{equation*}
\begin{aligned}
|\mathcal{L}G(v^{\varepsilon}_\ell)-{\bm Q} G(v^{\varepsilon}_\ell)|
=&\Big|\frac{1}{2}(K^\varepsilon)^2\sigma^2(v^{\varepsilon}_\ell) G''(v^{\varepsilon}_l)+\Big((v^{\varepsilon}_\ell-V^\varepsilon_0)\widehat{R}+K^\varepsilon b(v^{\varepsilon}_\ell)\Big) G'(v^{\varepsilon}_\ell)\\
&-\left(q_{l,l-1}G(v^{\varepsilon}_{l-1})+q_{l,l}G(v^{\varepsilon}_\ell)
+q_{l,l+1}G(v^{\varepsilon}_{l+1})\right)\Big|.
\end{aligned}
\end{equation*}
Then we use \eqref{q-solve}, the triangle inequality, the Lipschitz continuity and the linear growth of $b$ and $\sigma$ to give
\begin{equation}
\begin{aligned}\label{ineq:L-Q}
&|\mathcal{L}G(v^{\varepsilon}_\ell)-{\bm Q}G(v^{\varepsilon}_\ell)|\\
\le&C\Bigg[\left| G''(v^{\varepsilon}_\ell)-\frac{2}{(h_{l+1}+h_{l})}\left(\frac{G(v^{\varepsilon}_{l+1})
-G(v^{\varepsilon}_\ell)}{h_{l+1}}
-\frac{G(v^{\varepsilon}_{l})-G(v^{\varepsilon}_{l-1})}{h_{l}}\right)\right|\\
&+\bigg| G'(v^{\varepsilon}_\ell)-\frac{1}{(h_{l+1}+h_{l})}\left(\frac{h_{l}(G(v^{\varepsilon}_{l+1})
-G(v^{\varepsilon}_\ell))}{h_{l+1}}
+\frac{h_{l+1}(G(v^{\varepsilon}_{l})-G(v^{\varepsilon}_{l-1}))}{h_{l}}\right)\bigg|\Bigg].
\end{aligned}
\end{equation}
By Taylor expansion, we have that
 \begin{equation}
 \begin{aligned}\label{taylor}
 G(v^{\varepsilon}_{l+1})&=G(v^{\varepsilon}_\ell)
 +G^{\prime}(v^{\varepsilon}_\ell)h_{l+1}
 +G^{\prime\prime}(v^{\varepsilon}_{l})\frac{h^2_{l+1}}{2}+o(h^2_{l+1}),\\
 G(v^{\varepsilon}_{l-1})&=G(v^{\varepsilon}_\ell)
 -G^{\prime}(v^{\varepsilon}_\ell)h_{l}
 +G^{\prime\prime}(v^{\varepsilon}_\ell)\frac{h^2_{l}}{2}
 +o(h_{l}^2).
 \end{aligned}
 \end{equation}
 Plugging \eqref{taylor} into \eqref{ineq:L-Q} and using the condition $h_\ell=\mathcal{O}(M^{-1})$, $|h_\ell-h_{\ell+1}|\le \mathcal{O}(M^{-2})$, we obtain
 \begin{equation*}
|\mathcal{L}G(v^{\varepsilon}_\ell)-{\bm Q}^\gamma G(v^{\varepsilon}_\ell)|\le o(1).
\end{equation*}
Then by Lemma \ref{lem:ctmc-semi}, $\widetilde{V}^{\varepsilon}_{t}$ converges weakly to ${V}^{\varepsilon}_{t}$ as $M$ goes to infinity. This completes the proof.
\eproof

\vspace{0.3cm}

The next theorem shows the weak convergence  of $\widetilde X^\varepsilon_t$  to $X^\varepsilon_t$.
\begin{thm}\label{thm:con-ctmc-v}
Assuming that Assumption \ref{assum:mu} holds, the grid interval $h$, $\delta$ satisfy: $h_\ell=\mathcal{O}(M^{-1})$, $|h_\ell-h_{\ell+1}|\le \mathcal{O}(M^{-2})$, and $\delta_i=\mathcal{O}(N^{-1})$, $|\delta_i-\delta_{i+1}|\le\mathcal{O}(N^{-2})$, the continuous-time Markov chain $\widetilde{X}^{\varepsilon}_{t}$ converges weakly to ${X}^{\varepsilon}_{t}$, as $M$, $N$ go to infinity.
\end{thm}
\proof
It is easy to verify that under Assumption \ref{assum:mu}, $\theta(x^{\varepsilon},v^{\varepsilon})$, which is the drift term of ${X}^{\varepsilon}_{t}$ and defined by \eqref{def-theta},  is continuous and bounded for $(x^{\varepsilon},v^{\varepsilon})\in \mathbb X^\varepsilon\times\mathbb V^\varepsilon$. Then by similar arguments as
 Theorem \ref{thm:con-ctmc-vgamma}, we can obtain the desired conclusion. This completes the proof.
\eproof

\section{Option pricing under RSLV model}\label{sec:option}
In this section, we use the semimartingale and CTMC approximation techniques to give the explicit approximate  expression of option prices under the RSLV model.

Vanilla option prices for $(S_T,V_T)$ under the RSLV model \eqref{model:RSLV} is given by:
\begin{equation*}
\E\left[e^{-rT}\phi(S_T,V_T)|S_0=s,V_0=v\right],
\end{equation*}
where
\begin{equation*}
\phi(S_T,V_T)=\left\{
\begin{aligned}
&(S_T-D)^+~~~~for~a~call,\\
&(D-S_T)^+~~~~for~a~put,
\end{aligned}
\right.
\end{equation*}
and $r$ is the risk-free interest rate, $D$ is the strike price, and $T$ is the maturity. After the semimartingale approximation of $(S_T,V_T)$ by $(S^\varepsilon_T,V^\varepsilon_T)$, and the CTMC approximation by $(\widetilde S^\varepsilon_T,\widetilde V^\varepsilon_T)$, the European options prices under the RSLV model have the following approximate formula.
\begin{alg}[European options] \label{alg:European}
Given that $s=s^\varepsilon$, $v=v^{\varepsilon}=v^{\varepsilon}_l$ and $x^{\varepsilon}_i=g(s^\varepsilon)+\rho f(v^{\varepsilon}_l)$, the European option price under the RSLV model can be  approximately calculated by
\begin{equation}\label{ctmc-form-European}
\begin{aligned}
&\E\left[e^{-rT}\phi(S_T,V_T)|S_0=s,V_0=v\right]\\
\approx&\E\left[e^{-rT}\phi(S^\varepsilon_T,V^\varepsilon_T)|S^\varepsilon_0=s^\varepsilon,V^\varepsilon_0=v^\varepsilon\right]\\
\approx&\E\left[e^{-rT}\phi(g^{-1}\big(\widetilde X^\varepsilon_t+\rho f(\widetilde V^\varepsilon_t)\big),\widetilde V^\varepsilon_T)|\widetilde X^\varepsilon_0=x^\varepsilon_i,\widetilde V^\varepsilon_0=v^\varepsilon_l\right]\\
=&e^{-rT}\cdot{\bf e}_{i,l}\cdot\exp({\bf \Lambda}T)\cdot {\bf \Phi^{(1)}}.\nonumber
\end{aligned}
\end{equation}
Here ${\bf e}_{i,l}$ is a $1\times N M$ vector with all entries equal to $0$ except that the $(l-1)N+i$ entry is equal to $1$, ${\bf \Lambda}$ is given by \eqref{matrix}, and ${\bf \Phi^{(1)}}$ is an $N M \times 1$ vector with elements for $l\in\mathcal{M}$, $i\in\mathcal{N}$,
\begin{equation*}
{\bf \Phi^{(1)}}_{(l-1)N+i}=\phi\big(g^{-1}( x^\varepsilon_i + \rho f(v^\varepsilon_{l}))\big)=\left\{
\begin{aligned}
&\left(g^{-1}( x^\varepsilon_i + \rho f(v^\varepsilon_{l})) - D\right)^+~~~~for~a~call,\\
&\left(D-g^{-1}( x^\varepsilon_i + \rho f(v^\varepsilon_{l}))\right)^+~~~~for~a~put.
\end{aligned}
\right.
\end{equation*}
\end{alg}

Similarly, we have the semimartingale and CTMC approximate formula of  barrier options prices under the RSLV model.
\begin{alg}[barrier options] \label{alg:barrier}
Given that $s=s^\varepsilon$, $v=v^{\varepsilon}=v^{\varepsilon}_l$ and $x^{\varepsilon}_i=g(s^\varepsilon)+\rho f(v^{\varepsilon}_l)$, the barrier option price under the RSLV model with $0\le L<U<\infty$, can be  approximately calculated by
\begin{equation}\label{ctmc-form-barrier}
\begin{aligned}
&\E\left[e^{-rT}\phi(S_T,V_T)\1_{\{L<S_T<U\}}|S_0=s,V_0=v\right]\\
\approx&\E\left[e^{-rT}\phi(S^\varepsilon_T,V^\varepsilon_T)\1_{\{L<S_T<U\}}|S^\varepsilon_0=s^\varepsilon,V^\varepsilon_0=v^\varepsilon\right]\\
\approx&\E\left[e^{-rT}\phi(g^{-1}\big(\widetilde X^\varepsilon_T+\rho f(\widetilde V^\varepsilon_T)\big),\widetilde V^\varepsilon_T)\1_{\{L<g^{-1}\big(\widetilde X^\varepsilon_T+\rho f(\widetilde V^\varepsilon_T)\big)<U\}}|\widetilde X^\varepsilon_0=x^\varepsilon_i,\widetilde V^\varepsilon_0=v^\varepsilon_l\right]\\
=&e^{-rT}\cdot{\bf e}_{i,l}\cdot\exp({\bf \Lambda}T)\cdot {\bf \Phi^{(2)}}.\nonumber
\end{aligned}
\end{equation}
Here ${\bf e}_{i,l}$ is a $1\times N M$ vector with all entries equal to $0$ except that the $(l-1)N+i$ entry is equal to $1$, ${\bf \Lambda}$ is given by \eqref{matrix}, and ${\bf \Phi^{(2)}}$ is an $N M \times 1$ vector with elements for $l\in\mathcal{M}$, $i\in\mathcal{N}$,
\begin{equation*}
{\bf \Phi^{(2)}}_{(l-1)N+i}=
\left\{
\begin{aligned}
&\left(g^{-1}( x^\varepsilon_i + \rho f(v^\varepsilon_{l})) - D\right)^+\1_{\{L<g^{-1}\big(x^\varepsilon_i+\rho f(v^\varepsilon_l)\big)<U\}}~~~~for~a~call,\\
&\left(D-g^{-1}( x^\varepsilon_i + \rho f(v^\varepsilon_{l}))\right)^+\1_{\{L<g^{-1}\big(x^\varepsilon_i+\rho f(v^\varepsilon_l)\big)<U\}}~~~~for~a~put.
\end{aligned}
\right.
\end{equation*}
\end{alg}

Next we consider (finite-maturity) American options, whose prices are given by
\begin{equation*}
\max_{\tau\in\mathcal{T}}\E\left[e^{-r\tau}\phi(S_\tau,V_\tau)|S_0=s,V_0=v\right],
\end{equation*}
where the set $\mathcal{T}$ comprises of the collection of $\mathbb{F}$-stopping times taking values between $0$ and $T$. This means that American options can be exercised at any time in $[0,T]$. $\mathcal{T}$ can be approximated by a finite set of admissible exercise times $\mathcal{T}^{n}:=\{\tau_i\}^n_{i=0}$, where $\tau_i=\frac{iT}{n}$, where $n$ is the number of monitoring dates. Under admissible exercise times set $\mathcal{T}^{n}$, the option is called the Bermudan option. By semimartingale and CTMC approximations, the value of the American option under the RSLV model can be approximately expressed as:
\begin{alg}[American options] \label{alg:American}
Given that $s=s^\varepsilon$, $v=v^{\varepsilon}=v^{\varepsilon}_l$ and $x^{\varepsilon}_i=g(s^\varepsilon)+\rho f(v^{\varepsilon}_l)$, the Bermudan option price under RSLV model can be  approximately calculated by
\begin{equation}\label{ctmc-form-bermudan}
\begin{aligned}
&\max_{\tau\in\mathcal{T}}\E\left[e^{-r\tau}\phi(S_\tau,V_\tau)|S_0=s,V_0=v\right]\\
\approx&\max_{\tau\in\mathcal{T}^{n}}\E\left[e^{-r\tau}\phi(S_\tau,V_\tau)|S_0=s,V_0=v\right]\\
\approx&\max_{\tau\in\mathcal{T}^n}\E\left[e^{-r\tau}\phi(S^\varepsilon_\tau,V^\varepsilon_\tau)|S^\varepsilon_0=s^\varepsilon,V^\varepsilon_0=v^\varepsilon\right]\\
\approx&\max_{\tau\in\mathcal{T}^n}\E\left[e^{-r\tau}\phi(g^{-1}\big(\widetilde X^\varepsilon_\tau+\rho f(\widetilde V^\varepsilon_\tau)\big),\widetilde V^\varepsilon_\tau)|\widetilde X^\varepsilon_0=x^\varepsilon_i,\widetilde V^\varepsilon_0=v^\varepsilon_l\right]=:B_0,\nonumber
\end{aligned}
\end{equation}
where $B_0={\bf e}_{i,l}\cdot{\bf B_0}$, and
\begin{equation*}
\left\{
\begin{aligned}
{\bf B_n}&={\bf \Phi^{(1)}},\\
{\bf B_i}&=\max \left\{e^{-rT/n}\exp({\bf \Lambda}T/n)\cdot{\bf B_{i+1}},{\bf \Phi^{(1)}}\right\},~i=n-1,n-2,\dots,0.
\end{aligned}
\right.
\end{equation*}
Here $\bf{B_i}$ is a $N M\times1$ vector,  ${\bf e}_{i,l}$ a $1\times N M$ vector with all entries equal to $0$ except that the $(l-1)N+i$ entry is equal to $1$, ${\bf \Lambda}$ is given by \eqref{matrix}, and ${\bf \Phi^{(1)}}$ is an $N M \times 1$ vector with elements for $l\in\mathcal{M}$, $i\in\mathcal{N}$,
\begin{equation*}
{\bf \Phi^{(1)}}_{(l-1)N+i}=\phi\big(g^{-1}( x^\varepsilon_i + \rho f(v^\varepsilon_{l}))\big)=\left\{
\begin{aligned}
&\left(g^{-1}( x^\varepsilon_i + \rho f(v^\varepsilon_{l})) - D\right)^+~~~~for~a~call,\\
&\left(D-g^{-1}( x^\varepsilon_i + \rho f(v^\varepsilon_{l}))\right)^+~~~~for~a~put.
\end{aligned}
\right.
\end{equation*}
\end{alg}

\begin{remark}[Convergence analysis of options pricing]\label{re:con}
Note that Theorem \ref{thm:conver_s} and Theorem \ref{thm:con-ctmc-v} show respectively the weak convergence of the semimartingale approximation and the CTMC approximation. For European and barrier options, since the functions $\phi$, $g$, $f$  are all continuous, the options prices obtained by the semimartingale and CTMC approximations converge to the original RSLV options prices, due to the continuous mapping theorem (see, e.g., \citet{kushner}). For American options, the weak convergence follows from the following three types of convergence: the convergence of time step discretization (refer to Chapter $2.4$ in \citet{quecke}),  semimartingale approximation (due to continuous mapping theorem), and the CTMC approximation (see Section 6 in \citet{eriksson}).
\end{remark}

\section{Numerical results}\label{sec:numerical}

In this section, we extend the traditional stochastic local volatility model (see Table \ref{tab-slv-model}) to rough stochastic local volatility model (see Table \ref{tab-rslv-model}). Then a series of examples for European, barrier and American options are used to illustrate the accuracy and efficiency of the semimartingale and CTMC approximation method introduced in Section \ref{sec:option}.  All numerical experiments are carried out with Matlab R$2016$a on a Core i$7$ desktop with $16$GB RAM and speed $3.60$ GHz.

Table \ref{tab-slv-model} collects some popular stochastic local volatility models in the literature.  The traditional stochastic local volatility models are all driven by standard Brownian motions. However, as confirmed by \citet{gatheral}, for a very wide range of assets, historical fluctuation time series exhibit rougher behavior than the Brownian motion. Therefore, it is very practical to recast the traditional stochastic local volatility models in the rough setting. Similar to the way \citet{euch} deducing the rough Heston model, we generalize all models listed in Table \ref{tab-slv-model} to their rough counterparts. The results are listed in Table \ref{tab-rslv-model}. In order to numerically solve the option pricing problem under RSLV models, we use the semimartingale approximation introduced in Section \ref{sec:semi-approx} and auxiliary processes introduced in Lemma \ref{lem:AuxLem}  to approximately transform those models into PSLV models with two independent Brownian motions. The results are listed in Table \ref{tab-pslv-x}.

\begin{table}[htbp]
\centering
\small
\caption{Examples of the stochastic local volatility models.}\label{tab-slv-model}
\smallskip
\begin{threeparttable}
\begin{tabular}{|l|l|l|}
\hline
Heston &
$dS_t=(r-q)S_tdt+S_t\sqrt{V_t}dW_t$ &$\eta,\vartheta\in{\mathbb R}$\\
(\citet{heston})&
$dV_t=\eta(\vartheta-V_t)dt+\sigma\sqrt{V_t}dB_t$ &$\sigma>0$
\\
\hline
$4/2$ model &
$dS_t=(r-q)S_tdt+S_t[a\sqrt{V_t}+b/\sqrt{V_t}]dW_t$ &$\eta,\vartheta,a,b\in{\mathbb R}$\\
(\citet{grasselli}) &
$dV_t=\eta(\vartheta-V_t)dt+\sigma\sqrt{V_t}dB_t$
& $\sigma>0$
\\
\hline
$\alpha$-Hyper&
$dS_t=(r-q)S_tdt+S_t\exp (V_t)dW_t$ &$\eta\in{\mathbb R}$\\
(\citet{da})  &
$dV_t=(\eta-\vartheta\exp (a V_t))dt+\sigma dB_t$  & $\vartheta,a,\sigma>0$
\\
\hline
SABR&
$dS_t=V_tS_t^{\beta}dW_t$ &$\beta\in[0,1)$\\
(\citet{hagan})  &
$dV_t=\sigma V_tdB_t$ &$\sigma>0$
\\
\hline
Heston-SABR&
$dS_t=(r-q)S_tdt+\sqrt{V_t}S_t^{\beta}dW_t$ &$\beta\in[0,1)$ \\
(\citet{van}) &
$dV_t=\eta(\vartheta-V_t)dt+\sigma \sqrt{V_t}dB_t$ &$\eta,\vartheta, \sigma>0$
\\
\hline
Quadratic SLV&
$dS_t=(r-q)S_tdt+\sqrt{V_t}(aS_t^{2}+bS_t+c)dW_t$ &$4ac>b^2$\\
(\citet{lipton}) &
$dV_t=\eta(\vartheta-V_t)dt+\sigma \sqrt{V_t}dB_t$  &$a,\eta,\vartheta,\sigma>0$
\\
\hline
\end{tabular}
\begin{tablenotes}
\item[*] Here $r$ is the risk-free interest rate, $q$ is the dividend  yield,  $V_t$ is the volatility and satisfies different stochastic differential equation for different models.
  \end{tablenotes}
    \end{threeparttable}
\end{table}

\begin{table}[htbp]
\centering
\small
\caption{Examples of the rough stochastic local volatility models.}\label{tab-rslv-model}
\smallskip
  \begin{threeparttable}
\begin{tabular}{|l|l|l|}
\hline
\multirow{2}*{Rough Heston} &
$dS_t=(r-q)S_tdt+S_t\sqrt{V_t}dW_t$ &$\eta,\vartheta\in{\mathbb R}$\\
&
$V_t=V_0+\int^t_0K(t,s)\left(\eta(\vartheta-V_s)ds+\sigma\sqrt{V_s}dB_s\right)$ &$\sigma>0$
\\
\hline
\multirow{2}*{Rough $4/2$ model} &
$dS_t=(r-q)S_tdt+S_t[a\sqrt{V_t}+b/\sqrt{V_t}]dW_t$ &$\eta,\vartheta,a,b\in{\mathbb R}$\\
 &
$V_t=V_0+\int^t_0K(t,s)\left(\eta(\vartheta-V_s)ds+\sigma\sqrt{V_s}dB_s\right)$
& $\sigma>0$
\\
\hline
\multirow{2}*{Rough $\alpha$-Hyper}&
$dS_t=(r-q)S_tdt+S_t\exp (V_t)dW_t$ &$\eta\in{\mathbb R}$\\
 &
$V_t=V_0+\int^t_0K(t,s)\left((\eta-\vartheta\exp (a V_s))ds+\sigma dB_s\right)$ & $\vartheta,a,\sigma>0$
\\
\hline
\multirow{2}*{Rough SABR}&
$dS_t=V_tS_t^{\beta}dW_t$ &$\beta\in[0,1)$\\
  &
$V_t=V_0+\int^t_0K(t,s)\left(\sigma V_sdB_s\right)$ &$\sigma>0$
\\
\hline
\multirow{2}*{Rough Heston-SABR}&
$dS_t=(r-q)S_tdt+\sqrt{V_t}S_t^{\beta}dW_t$ &$\beta\in[0,1)$ \\
 &
$V_t=V_0+\int^t_0K(t,s)\left(\eta(\vartheta-V_s)ds+\sigma \sqrt{V_s}dB_s\right)$ &$\eta,\vartheta, \sigma>0$
\\
\hline
\multirow{2}*{Rough quadratic SLV}&
$dS_t=(r-q)S_tdt+\sqrt{V_t}(aS_t^{2}+bS_t+c)dW_t$ &$4ac>b^2$\\
 &
$V_t=V_0+\int^t_0K(t,s)\left(\eta(\vartheta-V_s)ds+\sigma \sqrt{V_s}dB_s\right)$ &$a, \eta,\vartheta,\sigma>0$
\\
\hline
\end{tabular}
\begin{tablenotes}
\item[*] Here $r$ is the risk-free interest rate, $q$ is the dividend  yield,  $V_t$ is the volatility process with initial value $V_0>0$,  and $K(t,s)=(t-s)^{H-\frac{1}{2}}/\Gamma(H+1/2)$ is the fractional kernel with the Hurst parameter $H\in(0,1/2)$.
  \end{tablenotes}
    \end{threeparttable}
\end{table}

\begin{table}[htbp]
\centering
\small
\caption{Examples of dynamics and variance transforms for the perturbed stochastic local volatility models.}\label{tab-pslv-x}
\smallskip
  \begin{threeparttable}
\begin{tabular}{|l l|}
\hline
{Perturbed Heston}  &
$\left\{
 \begin{array}{ll}
 dX^\varepsilon_t=\left(r-q-\frac{V^\varepsilon_t}{2}-\frac{\rho\eta(\vartheta-V^\varepsilon_t)}{\sigma}
+\frac{\rho\chi(V_t^{\varepsilon,\gamma})}{K^\varepsilon \sigma}\right)dt
+\sqrt{1-\rho^2}\sqrt{V^\varepsilon_t}dB^{\bot}_t \\
dV^\varepsilon_t=-\chi(V_t^{\varepsilon,\gamma})dt+K^\varepsilon \left(\eta(\vartheta-V^\varepsilon_t)dt
+\sigma\sqrt{V^\varepsilon_t}dB_t\right) \\
dV_t^{\varepsilon,\gamma}=(-\gamma V_t^{\varepsilon,\gamma}+\eta(\vartheta-V^\varepsilon_t))dt
+\sigma\sqrt{V^\varepsilon_t}dB_t\\
X^\varepsilon_t=\log S^\varepsilon_t-\frac{\rho V^\varepsilon_t}{K^\varepsilon \sigma}
\end{array}
\right.$\\
\hline
{Perturbed $4/2$ model} &
$\left\{
 \begin{array}{ll}
 dX^\varepsilon_t=\left(r-q+\frac{\rho}{\sigma}
 (\frac{\chi(V_t^{\varepsilon,\gamma})}{K^\varepsilon }
 -\eta(\vartheta-V^\varepsilon_t))(a+\frac{b}{V^\varepsilon_t})
 -\frac{\rho K^\varepsilon b\sigma}{2V^\varepsilon_t}\right)dt\\
 ~~~~~ -\frac{1}{2}(a\sqrt{V^\varepsilon_t}+\frac{b}{\sqrt{V^\varepsilon_t}})^2dt
 +\sqrt{1-\rho^2}(a\sqrt{V^\varepsilon_t}+\frac{b}{\sqrt{V^\varepsilon_t}})dB^{\bot}_t \\
 dV^\varepsilon_t=-\chi(V_t^{\varepsilon,\gamma})dt+K^\varepsilon \left(\eta(\vartheta-V^\varepsilon_t)dt
 +\sigma\sqrt{V^\varepsilon_t}dB_t\right)\\
 dV_t^{\varepsilon,\gamma}=\left(-\gamma V_t^{\varepsilon,\gamma}+\eta(\vartheta-V^\varepsilon_t)\right)dt
+\sigma\sqrt{V^\varepsilon_t}dB_t\\
 X_t=\log S^\varepsilon_t-\rho\frac{aV^\varepsilon_t+b\log V^\varepsilon_t}{K^\varepsilon \sigma}
\end{array}
\right.$\\
\hline
{Perturbed $\alpha$-Hyper }   &
$\left\{
 \begin{array}{l}
 dX^\varepsilon_t=(r-q-\frac{\rho\exp(V^\varepsilon_t)}{\sigma}
 (\frac{\chi(V_t^{\varepsilon,\gamma})}{K^\varepsilon }-\eta+\vartheta\exp (a V^\varepsilon_t))dt  \\
 ~~~~~+(\frac{\rho K^\varepsilon \sigma\exp(V^\varepsilon_t)}{2}
 -\frac{\exp(2V^\varepsilon_t)}{2})dt
 +\sqrt{1-\rho^2} \exp(V^\varepsilon_t)dB^{\bot}_t\\
dV^\varepsilon_t=-\chi(V_t^{\varepsilon,\gamma})dt+K^\varepsilon \left((\eta-\vartheta\exp (a V^\varepsilon_t))dt+\sigma dB_t\right)\\
 dV_t^{\varepsilon,\gamma}=\left(-\gamma V_t^{\varepsilon,\gamma}+\eta-\vartheta\exp (a V^\varepsilon_t)\right)dt
+\sigma dB_t\\
 X^\varepsilon_t=\log S^\varepsilon_t-\frac{\rho \exp(V^\varepsilon_t)}{K^\varepsilon \sigma}
\end{array}
\right.$\\
\hline
{Perturbed SABR} &
$\left\{
 \begin{array}{ll}
 dX^\varepsilon_t=\left(-\frac{\beta v^2_t}{2(1-\beta)\left(X^\varepsilon_t+\frac{\rho V^\varepsilon_t}{K^\varepsilon \sigma}\right)}+\frac{\rho\chi(V_t^{\varepsilon,\gamma})}{K^\varepsilon \sigma}\right)dt+\sqrt{1-\rho^2} V^\varepsilon_tdB^{\bot}_t \\
dV^\varepsilon_t=-\chi(V_t^{\varepsilon,\gamma})dt+K^\varepsilon \sigma V^\varepsilon_tdB_t \\
 dV_t^{\varepsilon,\gamma}=-\gamma V_t^{\varepsilon,\gamma}dt
+\sigma V^\varepsilon_tdB_t\\
X^\varepsilon_t=\frac{(S^\varepsilon_t)^{1-\beta}}{1-\beta}-\frac{\rho V^\varepsilon_t}{K^\varepsilon \sigma}
\end{array}
\right.$\\
\hline
{Perturbed Heston-SABR}   &
$\left\{
 \begin{array}{ll}
 dX^\varepsilon_t=\left((r-q)(1-\beta)(X^\varepsilon_t+\frac{\rho V^\varepsilon_t}{K^\varepsilon \sigma})-\frac{\beta V^\varepsilon_t}{2(1-\beta)(X^\varepsilon_t+\frac{\rho V^\varepsilon_t}{K^\varepsilon \sigma})}\right)dt\\
 ~~~~~+\frac{\rho}{\sigma}\left(\frac{\chi(V_t^{\varepsilon,\gamma})}{K^\varepsilon }
 -\eta(\vartheta-V^\varepsilon_t)\right)dt
 +\sqrt{1-\rho^2}\sqrt{v_t}dB^{\bot}_t \\
dV^\varepsilon_t=-\chi(V_t^{\varepsilon,\gamma})dt+K^\varepsilon \left(\eta(\vartheta-V^\varepsilon_t)dt
+\sigma\sqrt{V^\varepsilon_t}dB_t\right) \\
dV_t^{\varepsilon,\gamma}=(-\gamma V_t^{\varepsilon,\gamma}+\eta(\vartheta-V^\varepsilon_t))dt
+\sigma\sqrt{V^\varepsilon_t}dB_t\\ X^\varepsilon_t=\frac{(S^\varepsilon_t)^{1-\beta}}{1-\beta}-\frac{\rho V^\varepsilon_t}{K^\varepsilon \sigma}
\end{array}
\right.$\\
\hline
{Perturbed quadratic SLV}  &
$\left\{
 \begin{array}{ll}
 dX^\varepsilon_t=\left(\frac{(r-q)S^\varepsilon_t}{a(S^\varepsilon_t)^2+bS^\varepsilon_t+c}
-\frac{2aS^\varepsilon_t+b}{2}V^\varepsilon_t+\frac{\rho}{\sigma}
(\frac{\chi(V_t^{\varepsilon,\gamma})}{K^\varepsilon }-\eta(\vartheta-V^\varepsilon_t))\right)dt\\
~~~~~+\sqrt{1-\rho^2}\sqrt{v_t}dB^{\bot}_t  \\
dV^\varepsilon_t=-\chi(V_t^{\varepsilon,\gamma})dt+K^\varepsilon \left(\eta(\vartheta-V^\varepsilon_t)dt
+\sigma\sqrt{V^\varepsilon_t}dB_t\right) \\
dV_t^{\varepsilon,\gamma}=(-\gamma V_t^{\varepsilon,\gamma}+\eta(\vartheta-V^\varepsilon_t))dt
+\sigma\sqrt{V^\varepsilon_t}dB_t\\ X^\varepsilon_t=\frac{2\arctan\left(\frac{2aS^\varepsilon_t+b}{\sqrt{4ac-b^2}}\right)}{\sqrt{4ac-b^2}}-\frac{\rho V^\varepsilon_t}{K^\varepsilon \sigma}
\end{array}
\right.$\\
\hline
\end{tabular}
\begin{tablenotes}
\item[*] Here $r$ is the risk-free interest rate, $q$ the dividend  yield, $X^\varepsilon_t$, $V^\varepsilon_t$, $V_t^{\varepsilon,\gamma}$ and $S^\varepsilon_t$ defined by \eqref{model:Aux}, $\chi(V_t^{\varepsilon,\gamma}):=\int^\infty_0e^{-\gamma\varepsilon}
\gamma V_t^{\varepsilon,\gamma}m(d\gamma)$, $K^\varepsilon =\varepsilon^{H-\frac{1}{2}}/\Gamma(H+1/2)$ the with the perturbation parameter $0<\varepsilon<<1$  and  the Hurst parameter  $H\in(0,1/2)$.
  \end{tablenotes}
    \end{threeparttable}
\end{table}

Next, we use the method introduced in Section \ref{sec:option} to numerically solve the European and barrier options pricing under RSLV models. All models considered in this paper share the same set of parameters:
\begin{equation*}
\begin{aligned}
&S_0=10,~ V_0=0.04,~ T=1,~ D=4,~ \rho=-0.75,~ r=0, ~
\sigma=0.8,~ \beta=0.7,\\
&\eta=4, ~\vartheta=0.035,~
a=0.02,~ b=0.05,~ c=1, ~L=2,~ U=15,~H=0.12.
\end{aligned}
\end{equation*}

In the numerical implementation, we use the same boundary points $v_{\min}=10^{-3}V_0$, $v_{\max}=4V_0$, $s_{\min}=10^{-3}S_0$, $s_{\max}=4S_0$ and $x_{\min}=10^{-3}X_0$, $x_{\max}=4X_0$, with $X_0=g(S_0)-\rho f(v_0)$ and choose the piecewise uniform grids.

\begin{example}[European and barrier options]
The value of European and barrier options can be calculated by Algorithm \ref{alg:European} and Algorithm \ref{alg:barrier}. Due to the independence between the auxiliary process $X^\varepsilon_t$ and the volatility process $V^\varepsilon_t$, the following fast algorithm can be used to speed up calculations.

\begin{alg}[A fast 2 dimensional CTMC algorithm for European and barrier options] \label{alg:fast}
Under the setting that $g(s^\varepsilon)-\rho f(v^\varepsilon)=x^\varepsilon=x^\varepsilon_i$, $v^\varepsilon=v^\varepsilon_\ell$,
\begin{equation}\label{ctmc-formula}
\begin{aligned}
&\E\left[ \phi(S^\varepsilon_T,V^\varepsilon_T)|S^\varepsilon_t=s^\varepsilon,V^\varepsilon_t=v^\varepsilon\right]\\
\approx& \E\left[ \phi(g^{-1}\big(\widetilde X^\varepsilon_t+\rho f(\widetilde V^\varepsilon_t)\big),\widetilde V^\varepsilon_T)|\widetilde X^\varepsilon_t=x^\varepsilon_i,\widetilde V^\varepsilon_t=v^\varepsilon_\ell\right]\\
=&\sum^M_{j=1}(e^{{\bm Q}T})_{\ell,j}\E\left[ \phi(g^{-1}\big(\widetilde X^\varepsilon_T+\rho f(\widetilde V^\varepsilon_T)\big),v^\varepsilon_j)|\widetilde X^\varepsilon_t=x^\varepsilon_i\right]\\
=&\sum^M_{j=1}(\exp({\bm Q}(T-t)))_{\ell,j}\times \mathbf{e}_{i} \cdot \exp (\mathbf{\Lambda}_l(T-t)) \cdot \mathbf{\Phi}_\ell,
\end{aligned}
\end{equation}
where ${\bf e}_{i}$ is a $1\times N$ vector with all entries equal to $0$ except that the $i$th entry is equal to $1$, ${\bm Q}$ is given by \eqref{q-solve}, ${\bf \Lambda}_\ell$ is given by \eqref{def-lambda}, and the payoff vector ${\bf \Phi}_l$ is a $N\times 1$ vectors with elements $({ \bf \Phi}_\ell)_{i}= \phi(g^{-1}\big(x^\varepsilon_i+\rho f(v^\varepsilon_\ell)\big),v^\varepsilon_\ell)$, for $\ell\in\mathcal{M}$, $i\in\mathcal{N}$. For European option, we have
\begin{equation*}
\phi\big(g^{-1}( x^\varepsilon_i + \rho f(v^\varepsilon_{l}))\big)=\left\{
\begin{aligned}
&\left(g^{-1}( x^\varepsilon_i + \rho f(v^\varepsilon_{l})) - D\right)^+~~~~for~a~call,\\
&\left(D-g^{-1}( x^\varepsilon_i + \rho f(v^\varepsilon_{l}))\right)^+~~~~for~a~put.
\end{aligned}
\right.
\end{equation*}
and for barrier option, there is
\begin{equation*}
\phi\big(g^{-1}( x^\varepsilon_i + \rho f(v^\varepsilon_{l}))\big)=
\left\{
\begin{aligned}
&\left(g^{-1}( x^\varepsilon_i + \rho f(v^\varepsilon_{l})) - D\right)^+\1_{\{L<g^{-1}\big(x^\varepsilon_i+\rho f(v^\varepsilon_l)\big)<U\}}~~~~for~a~call,\\
&\left(D-g^{-1}( x^\varepsilon_i + \rho f(v^\varepsilon_{l}))\right)^+\1_{\{L<g^{-1}\big(x^\varepsilon_i+\rho f(v^\varepsilon_l)\big)<U\}}~~~~for~a~put.
\end{aligned}
\right.
\end{equation*}
\end{alg}

\begin{remark}\label{re:ctmc}
When applying the CTMC method to a two-dimensional problem, the two-dimensional probability transition matrix is usually converted into a large matrix  (see \citet{song2016} and \citet{xi}). In this way, we need to calculate the matrix exponential where the exponent is a $NM \times NM$ matrix. This is time consuming, and requires a lot of computer storage space for intermediate inputs. Since the two stochastic processes are separated by Lemma \ref{lem:AuxLem}, we only need to compute $N\times N$ matrices $M$ times using our method. This greatly reduces the unnecessary calculation, reduces the computer storage space, and improves the operation efficiency. See Figure \ref{fig-time} for the comparison of actual operation efficiency.
\end{remark}

\begin{figure}[htbp]
\centering
 \subfigure[]{\label{subfig-time-euro}
 \includegraphics[width=6.5cm]{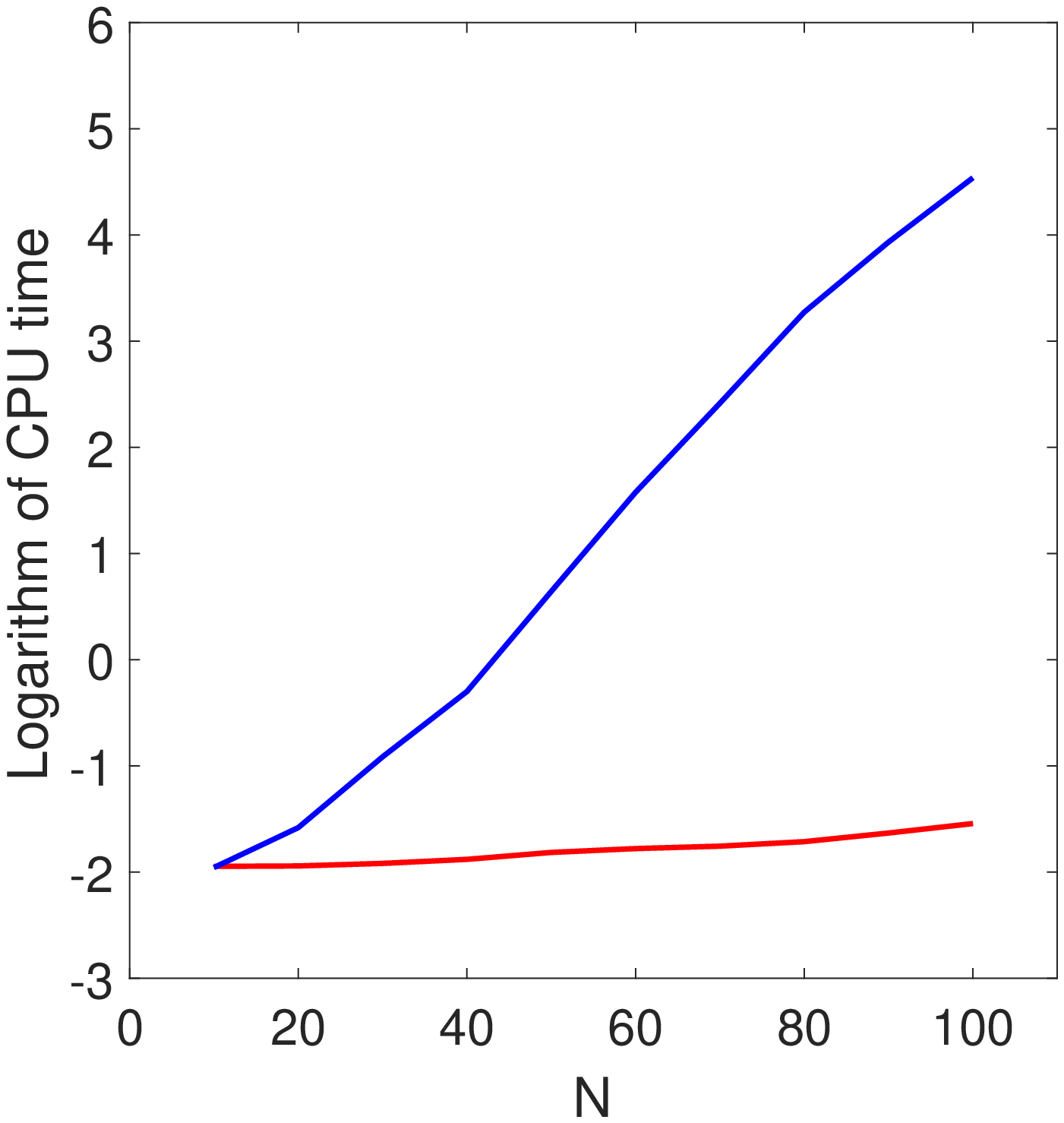}
}
 \subfigure[]{\label{subfig-time-barrier}
 \includegraphics[width=6.5cm]{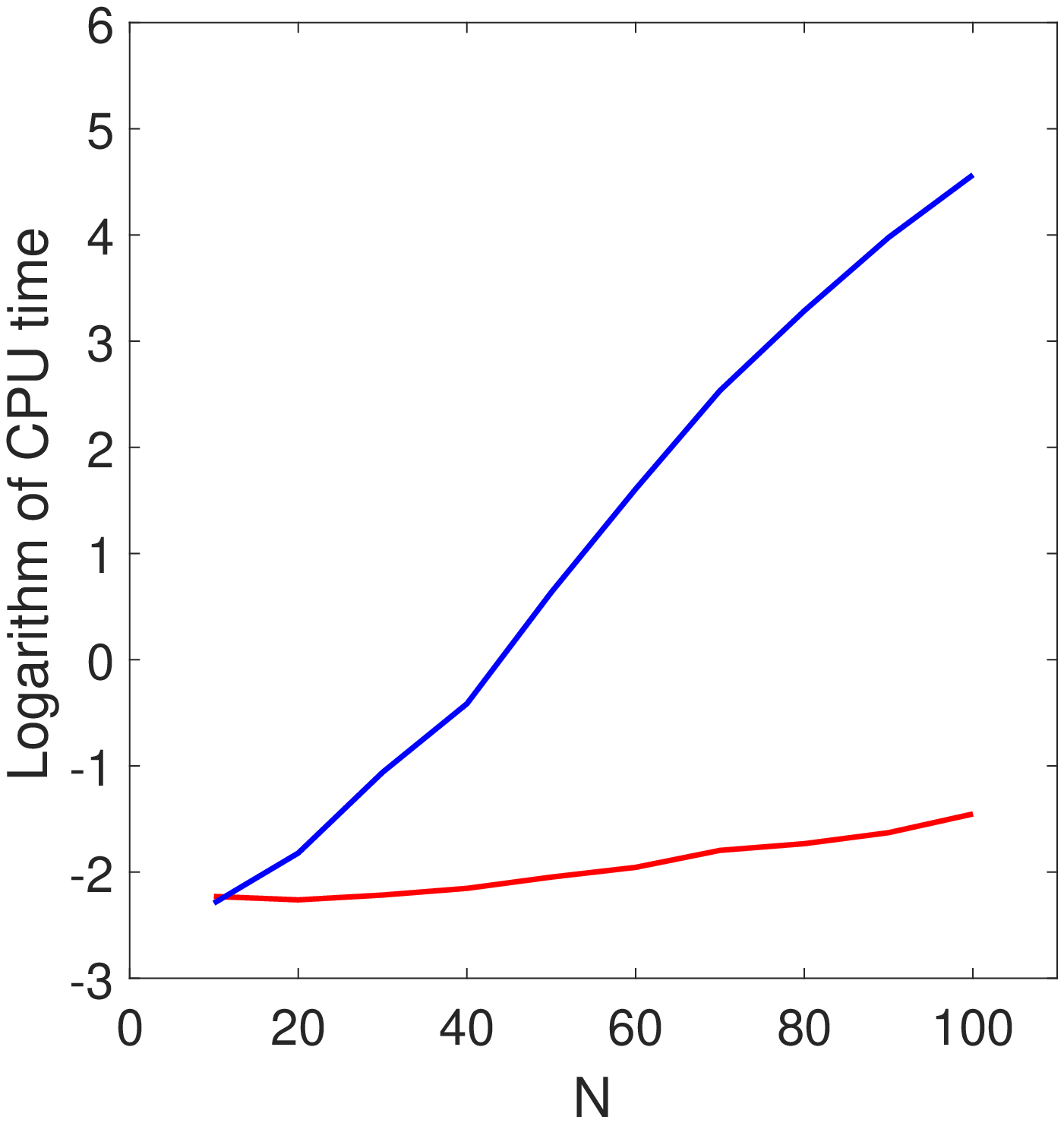}
}
 \caption{The plot of the number of grid points versus logarithmic of CPU times (seconds) for European (\ref{subfig-time-euro}) and barrier (\ref{subfig-time-barrier}) options under rough Heston model.  Here we set $M=N$. The blue lines in \ref{subfig-time-euro} and \ref{subfig-time-barrier} are obtained by Algorithm \ref{alg:European} and \ref{alg:barrier}, respectively.  The red lines are obtained by Algorithm \ref{alg:fast}. }\label{fig-time}
\end{figure}

\begin{table}[htbp]
\centering
\caption{European call option under the RSLV models.}\label{tab-european-epsilon}
\begin{threeparttable}
\begin{tabular}{l l l l l l l}
 \toprule
 \multirow{2}*{$\varepsilon$} & \multicolumn{5}{c}{Relative Errors}\\
\cline{2-7}
&R-H &R-$4/2$ &R-$\alpha$-H &R-SABR &R-H-SABR &R-Q-SLV\\
\hline
$10^{-4}$  &4.71e-3  &9.98e-3  &5.76e-3 &8.78e-3 &4.21e-3  & 5.86e-3\\
\hline
$10^{-5}$  &2.84e-3  &9.68e-3  &2.56e-3 &3.81e-3 &2.86e-3  & 4.53e-3\\
\hline
$10^{-6}$  &2.07e-3  &7.92e-3  &1.22e-3 &1.75e-3 &2.32e-3  & 3.97e-3\\
\hline
$10^{-7}$  &1.75e-3  &7.55e-3  &6.66e-4 &8.84e-4 &2.09e-3  & 3.44e-3\\
\hline
$10^{-8}$  &1.61e-3  &6.57e-3 &4.34e-4 &5.25e-4 &1.99e-3  & 2.98e-3\\
\hline
benchmark &6.0545  &0.0362  &6.0001 &4.9269 &6.0018 & 6.0000\\
 \bottomrule
\end{tabular}
\begin{tablenotes}
\item[*] Here ``R-H", ``R-$4/2$", ``R-$\alpha$-H", ``R-SABR", ``R-H-SABR", ``R-Q-SLV" represent ``rough Heston", ``rough $4/2$ model", ``rough $\alpha$-Hyper", ``rough SABR", ``rough Heston-SABR", ``rough quadratic SLV" models respectively. Benchmarks are obtained by fast simulation algorithm based on the Monte Carlo method in \citet{ma2021} with $10^5$ simple paths, and they take about $610.86$ seconds on average. The results in the table are calculated with $N=M=100$ via Algorithm \ref{alg:fast}, and  take only $0.18$ seconds on average.
    \end{tablenotes}
    \end{threeparttable}
\end{table}

\begin{figure}[htbp]
\centering
 \subfigure[]{\label{subfig-E-heston}
 \includegraphics[width=4.5cm]{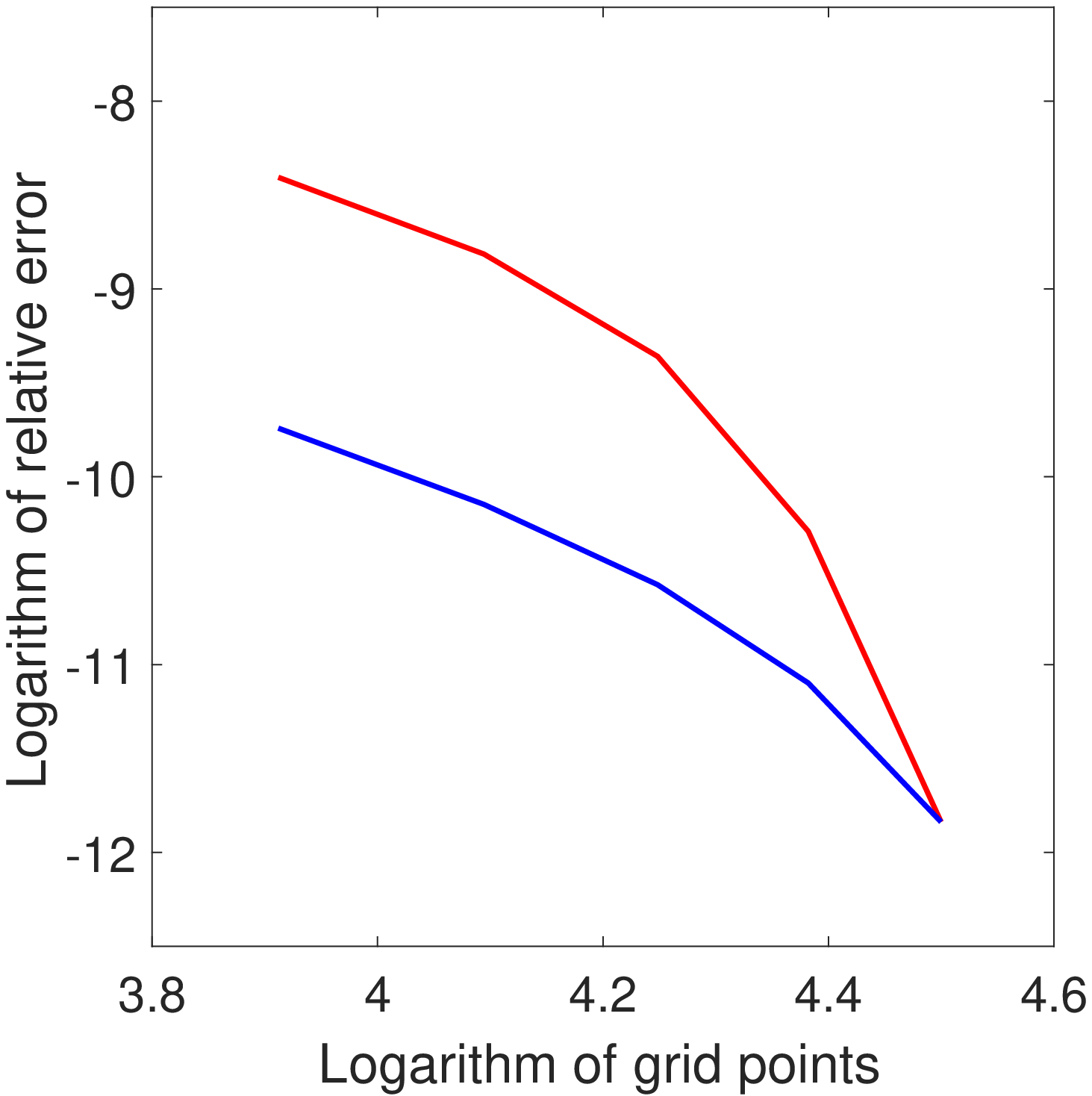}
}
 \subfigure[]{\label{subfig-E-fourtwo}
 \includegraphics[width=4.5cm]{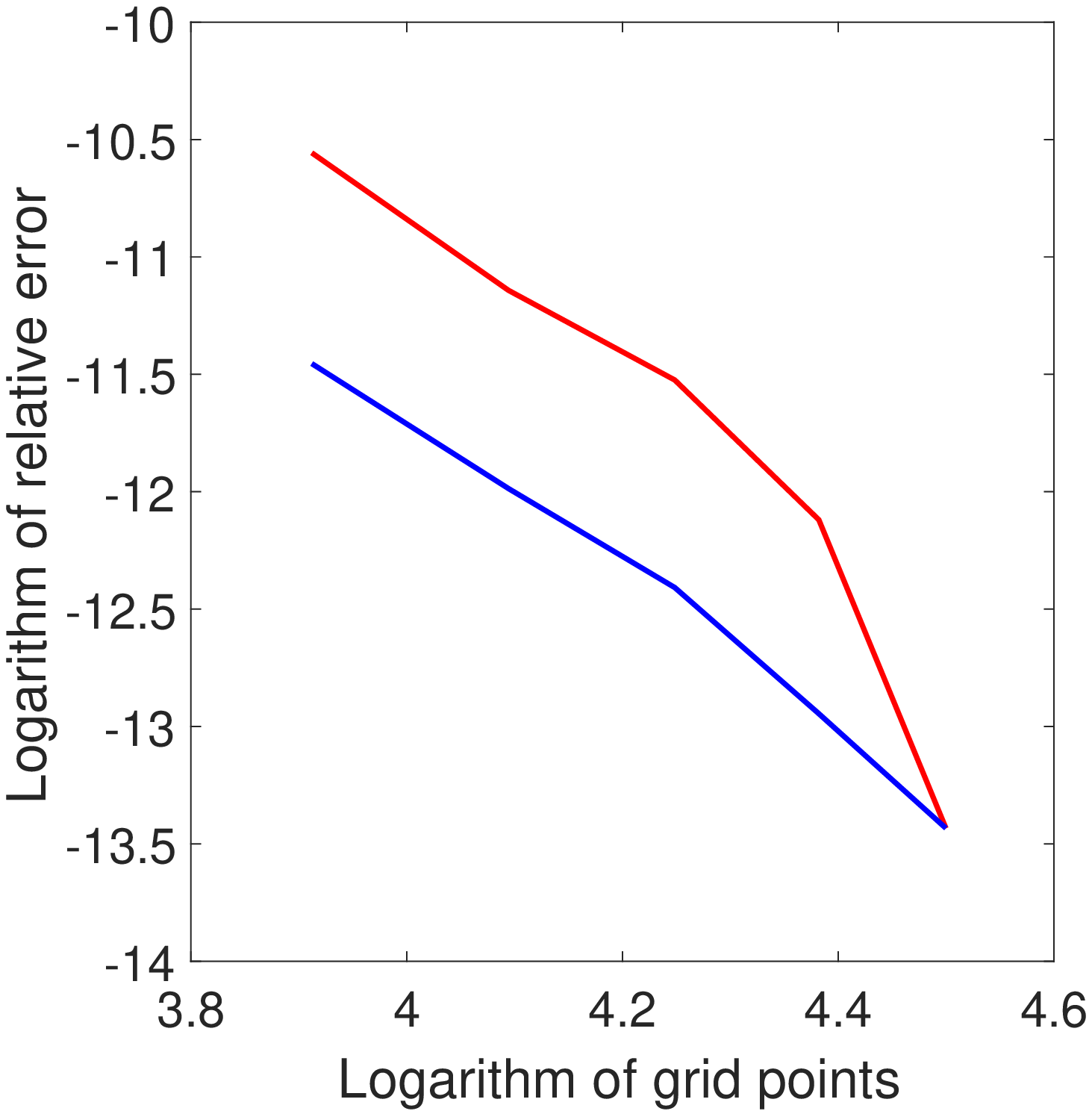}
}
 \subfigure[]{\label{subfig-E-alpha}
 \includegraphics[width=4.5cm]{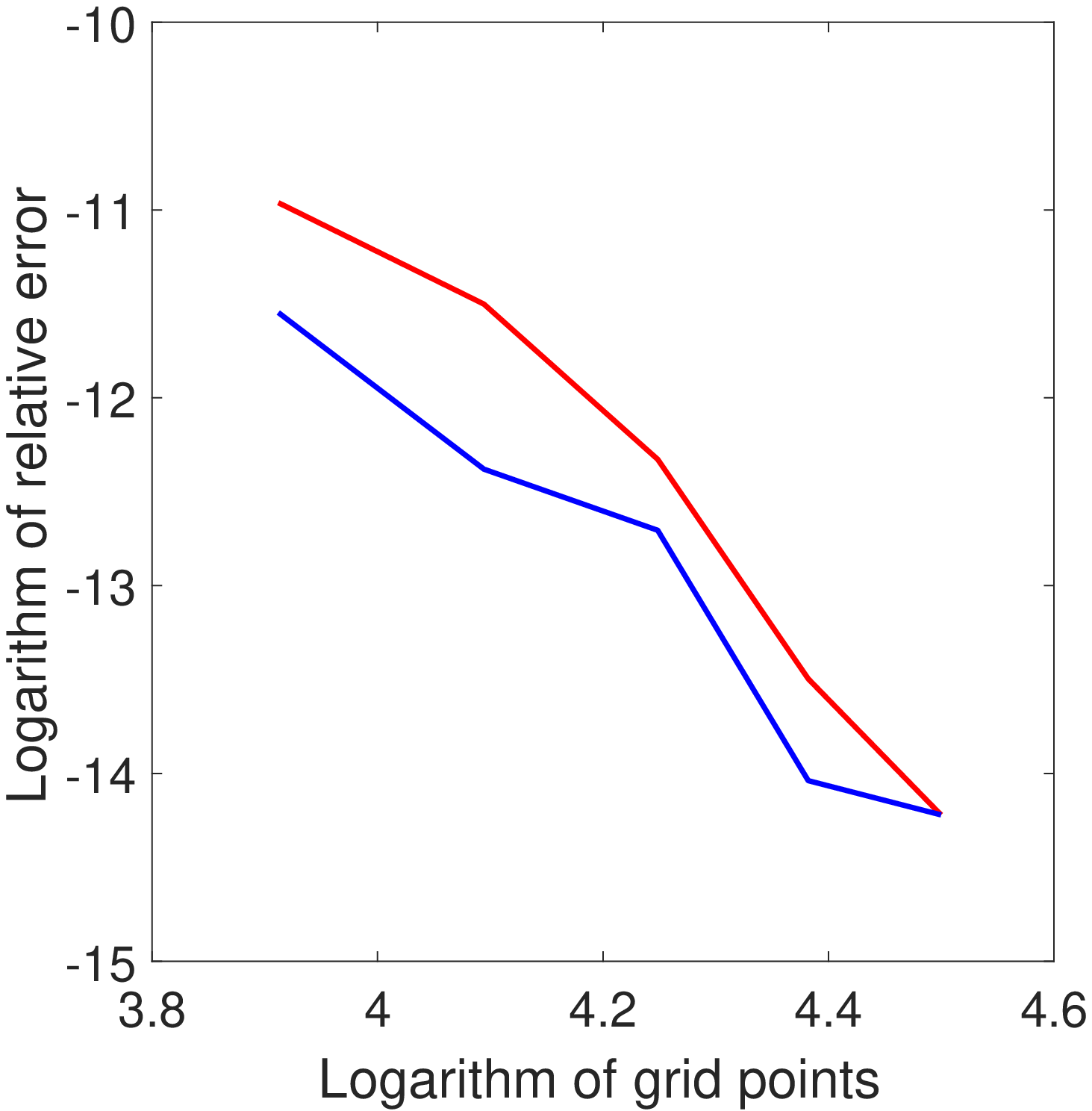}
}
 \subfigure[]{\label{subfig-E-sabr}
 \includegraphics[width=4.5cm]{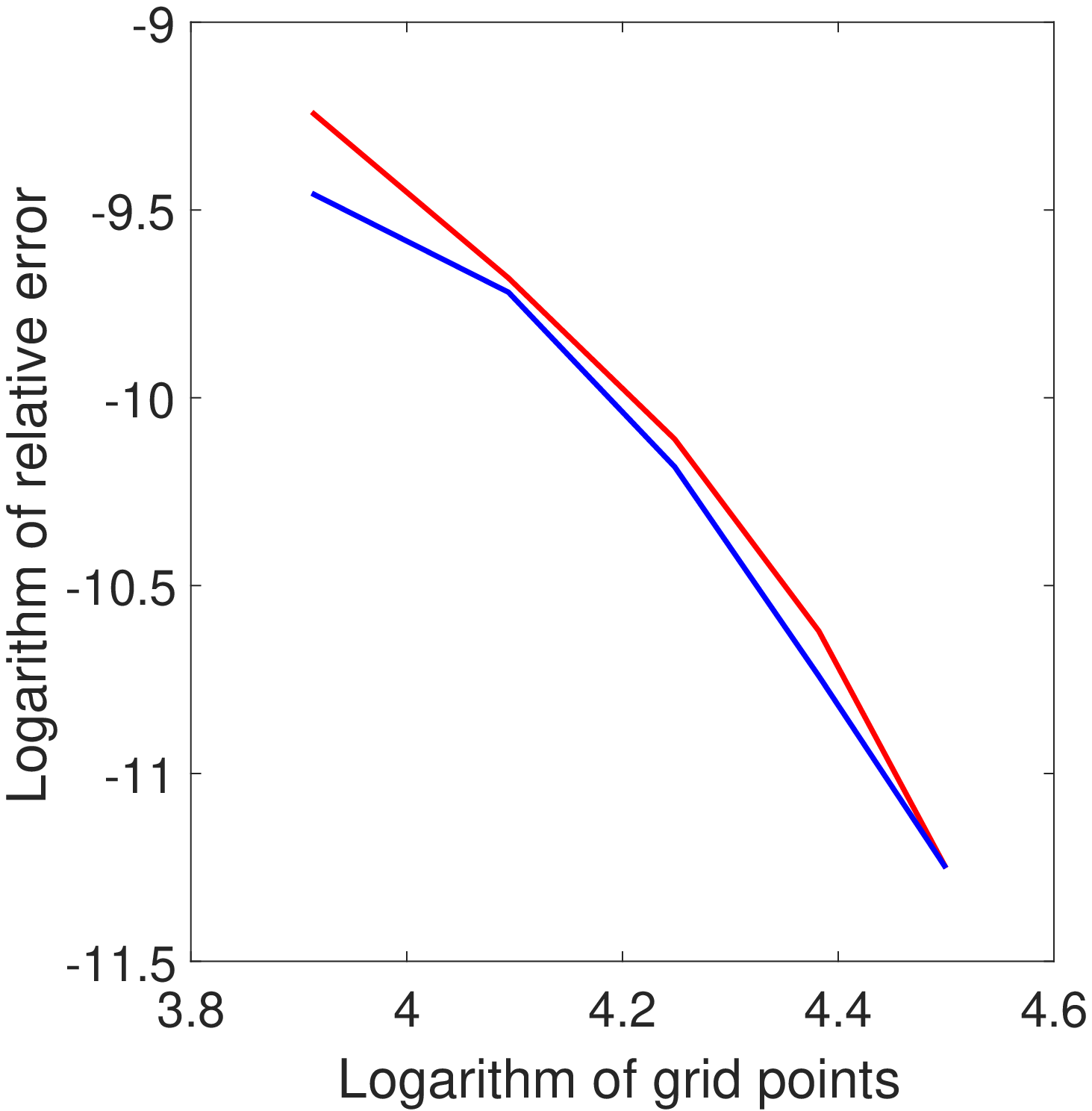}
}
 \subfigure[]{\label{subfig-E-hestonsabr}
 \includegraphics[width=4.5cm]{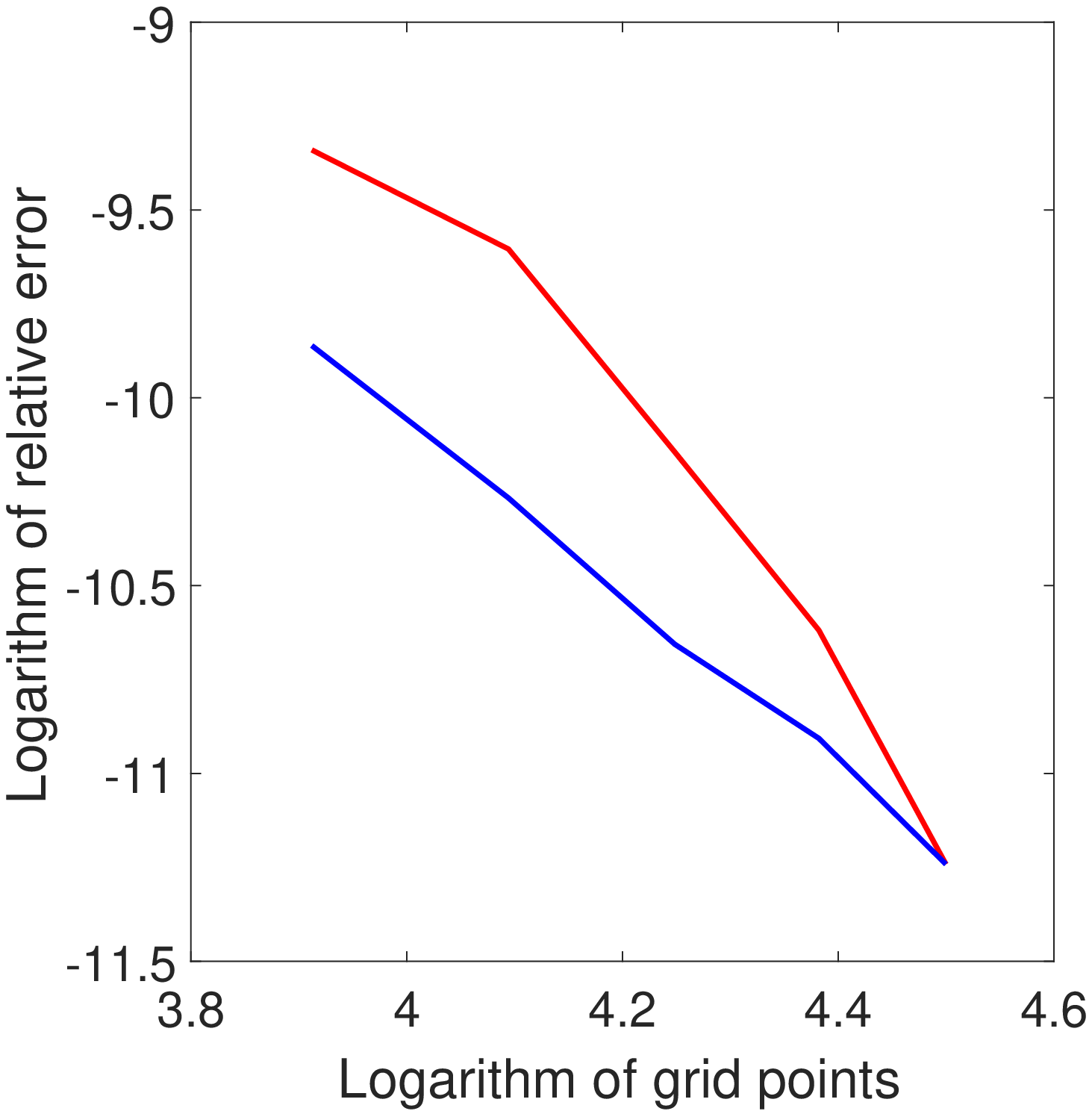}
}
 \subfigure[]{\label{subfig-E-qslv}
 \includegraphics[width=4.5cm]{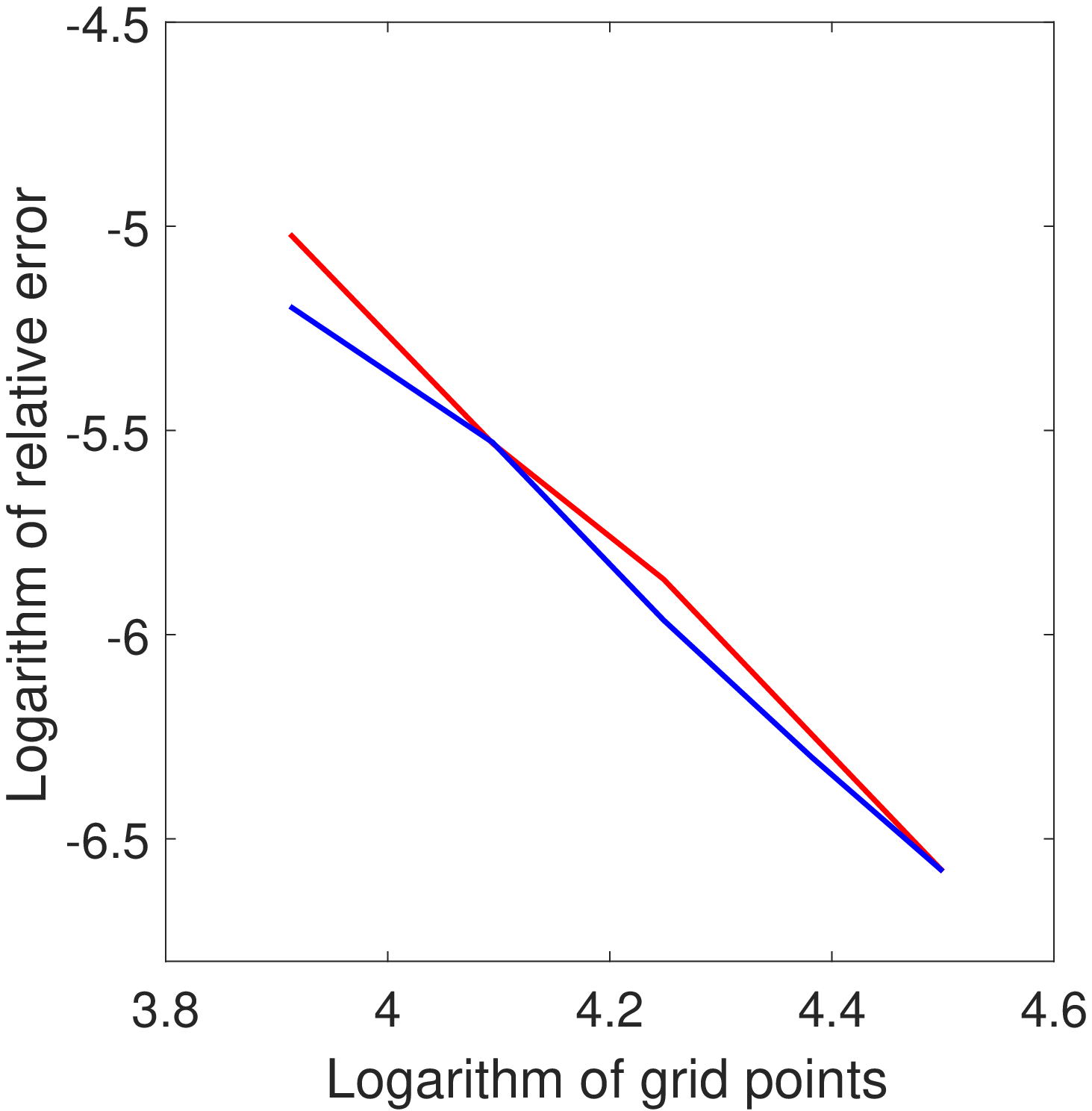}
}
 \caption{The logarithmic plot of the number of grid points versus relative error for European options under rough stochastic local volatility models via Algorithm \ref{alg:fast}.  Figures \ref{subfig-E-heston}, \ref{subfig-E-fourtwo}, \ref{subfig-E-alpha}, \ref{subfig-E-sabr}, \ref{subfig-E-hestonsabr}, \ref{subfig-E-qslv} correspond to rough Heston, $4/2$, $\alpha$-Hyper, SABR, Heston-SABR, quadratic SLV  models, respectively. The red lines are obtained by fixing $M=90$, and then increasing the grid points of $N$ from $50$ to $90$. The blue lines are obtained by fixing $N=90$ and changing $M$. Here we take the prices obtained by $M=N=100$, $\varepsilon=10^{-10}$ as the benchmarks to calculate relative errors. }\label{fig-European}
\end{figure}

\begin{table}[htbp]
\centering
\caption{Barrier call option under the RSLV models.}\label{tab-barrier-epsilon}
\begin{threeparttable}
\begin{tabular}{l l l l l l l}
 \toprule
 \multirow{2}*{$\varepsilon$} & \multicolumn{5}{c}{Relative Errors}\\
\cline{2-7}
&R-H &R-$4/2$ &R-$\alpha$-H &R-SABR &R-H-SABR &R-Q-SLV\\
\hline
$10^{-4}$  &4.21e-3  &7.52e-3  &3.54e-3 &1.51e-3 &2.84e-3  & 9.34e-3\\
\hline
$10^{-5}$  &2.86e-3  &6.18e-3  &1.68e-3 &7.42e-4 &2.07e-3  & 5.76e-3\\
\hline
$10^{-6}$  &2.32e-3  &5.63e-3  &9.07e-4 &4.21e-4 &1.75e-3  & 2.56e-3\\
\hline
$10^{-7}$  &2.08e-3  &5.40e-3  &5.86e-4 &2.88e-4 &1.61e-3  & 1.22e-3\\
\hline
$10^{-8}$  &1.99e-3  &5.30e-3  &4.53e-4 &2.32e-4 &1.56e-3  & 6.66e-4\\
\hline
benchmark &6.0492  &0.0345  &5.9753 &4.8099 &6.0000 & 5.9814\\
 \bottomrule
\end{tabular}
\begin{tablenotes}
\item[*] Here "R-H", "R-$4/2$", "R-$\alpha$-H", "R-SABR", "R-H-SABR", "R-Q-SLV" represent "rough Heston", "rough $4/2$ model", "rough $\alpha$-Hyper", "rough SABR", "rough Heston-SABR", "rough rough quadratic SLV" models respectively. Benchmarks are obtained by fast simulation algorithm based on Monte Carlo method in \citet{ma2021} with $10^5$ simple paths, and they take about $800.13$ seconds on average. The results in the table are calculated with $N=M=100$ via Algorithm \ref{alg:fast}, and they take only $0.20$ seconds on average.
    \end{tablenotes}
    \end{threeparttable}
\end{table}

\begin{figure}[htbp]
\centering
 \subfigure[]{\label{subfig-b-heston}
 \includegraphics[width=4.5cm]{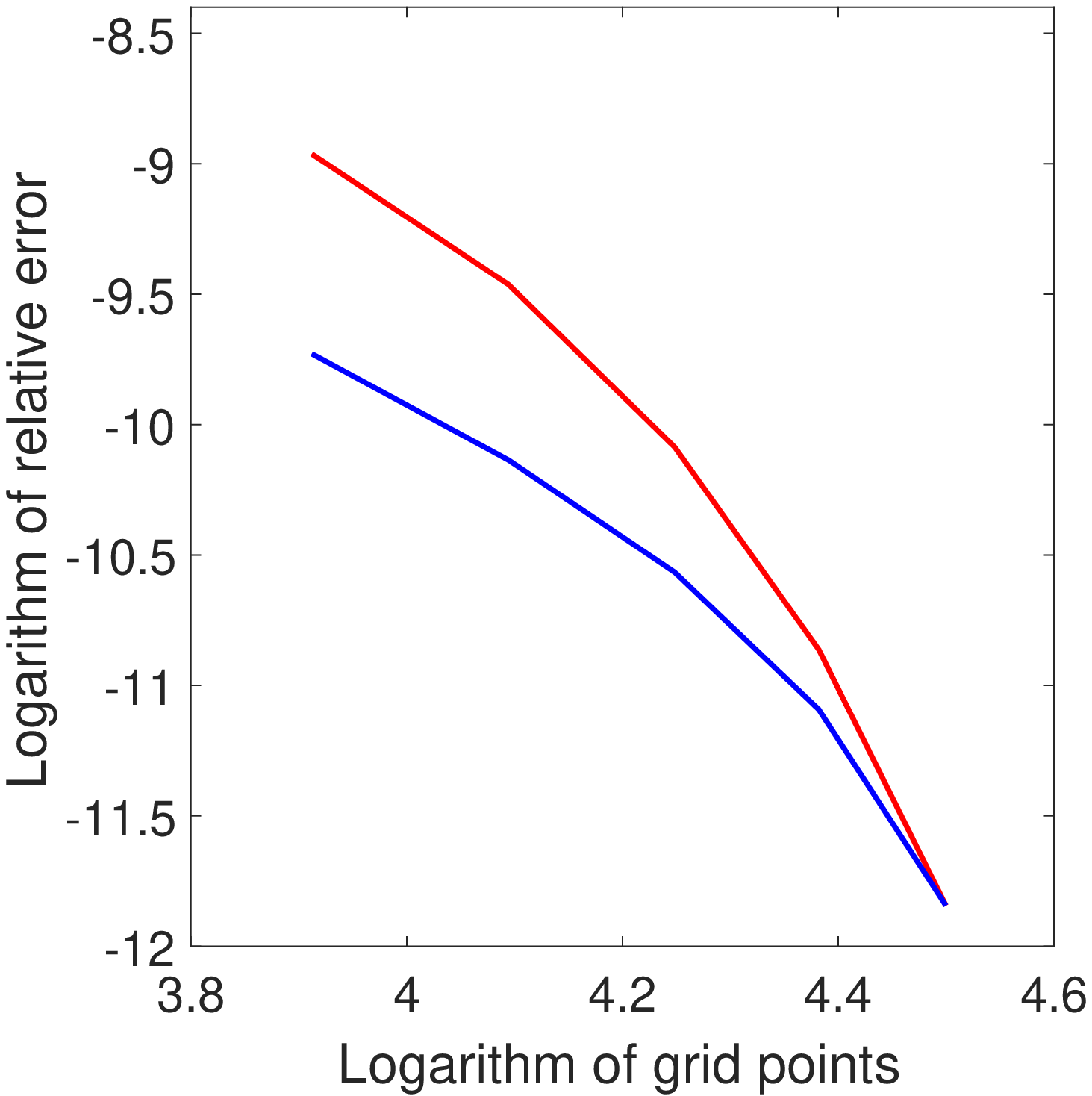}
}
 \subfigure[]{\label{subfig-b-fourtwo}
 \includegraphics[width=4.5cm]{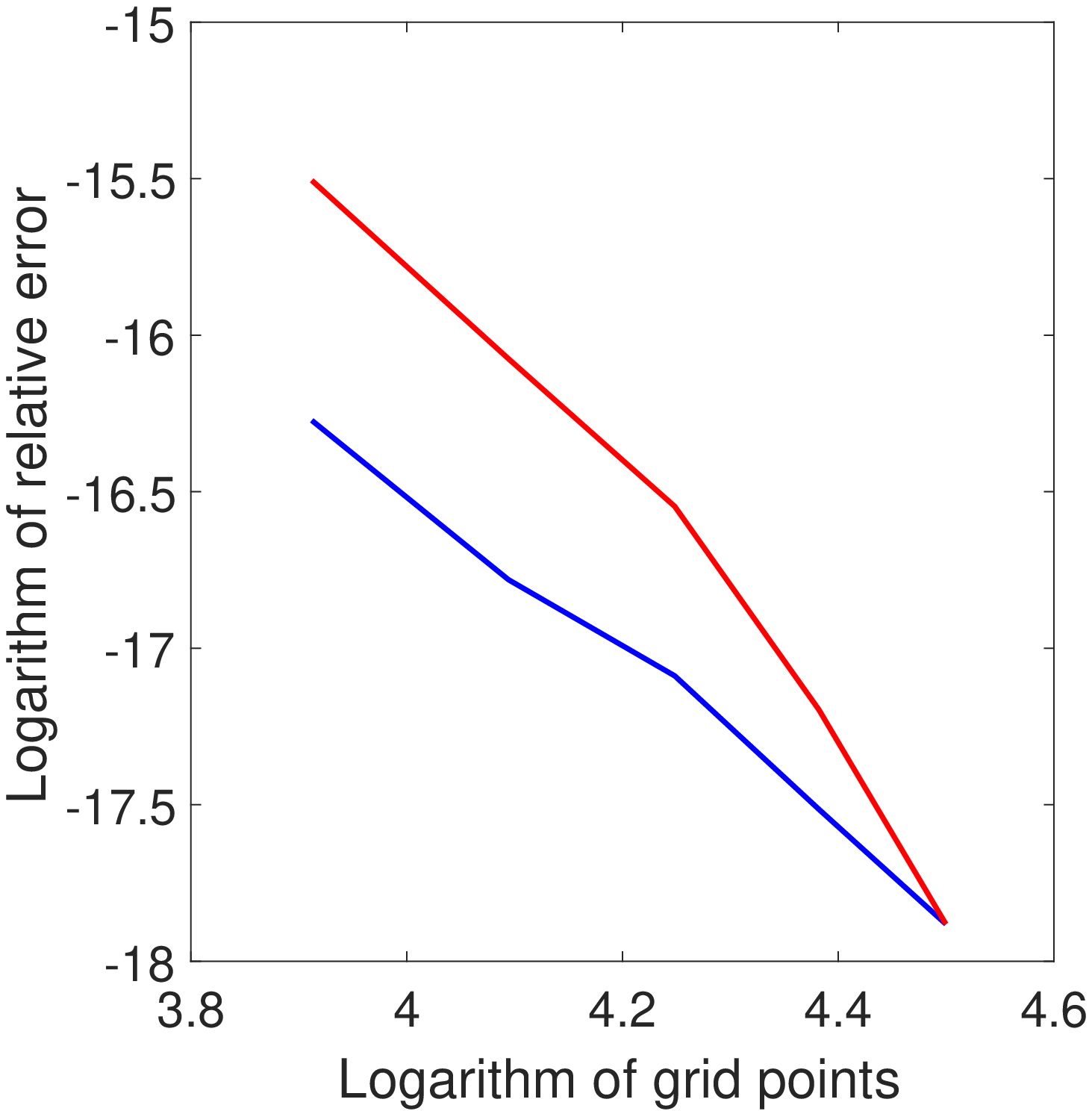}
}
 \subfigure[]{\label{subfig-b-alpha}
 \includegraphics[width=4.5cm]{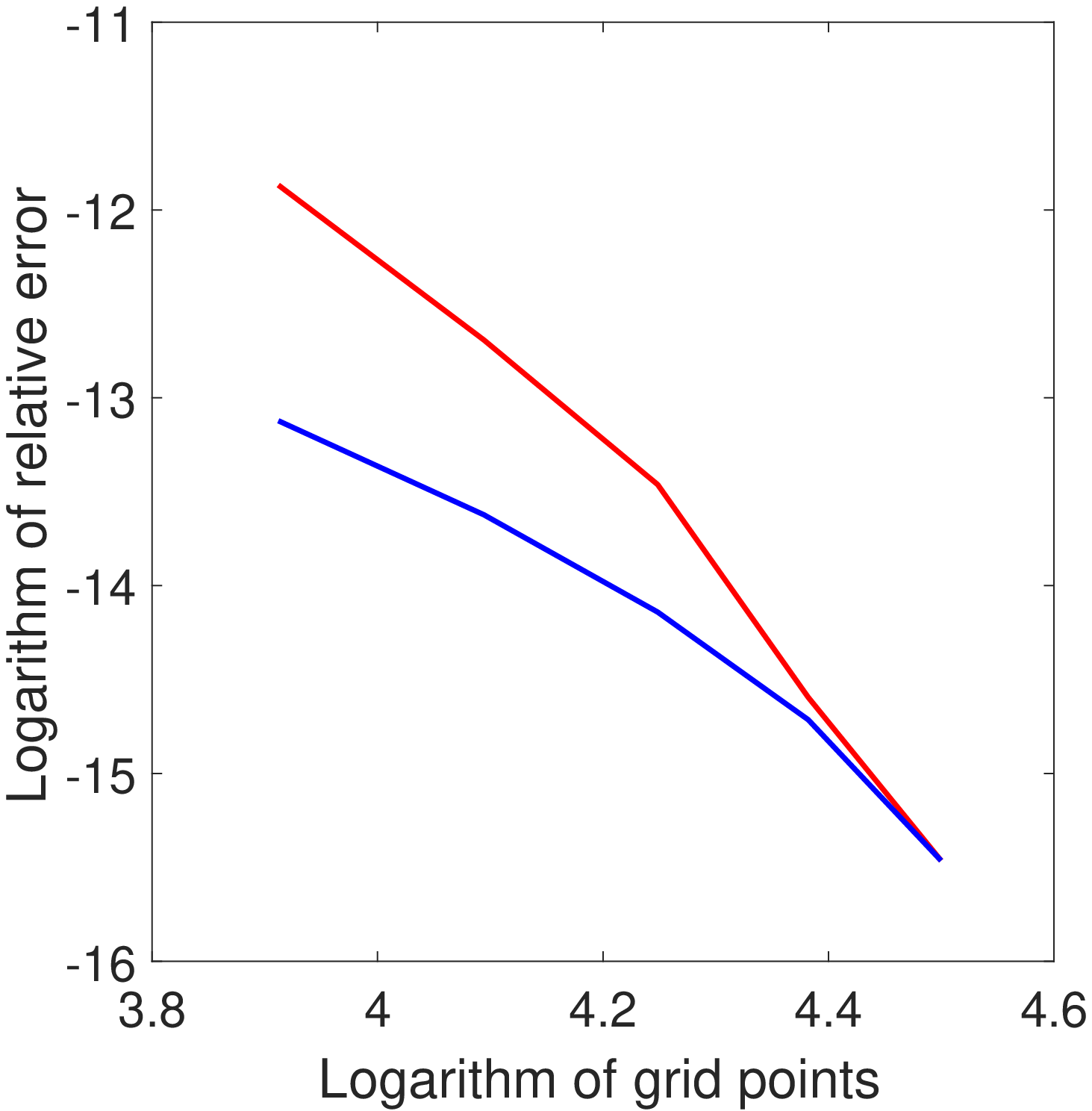}
}
 \subfigure[]{\label{subfig-b-sabr}
 \includegraphics[width=4.5cm]{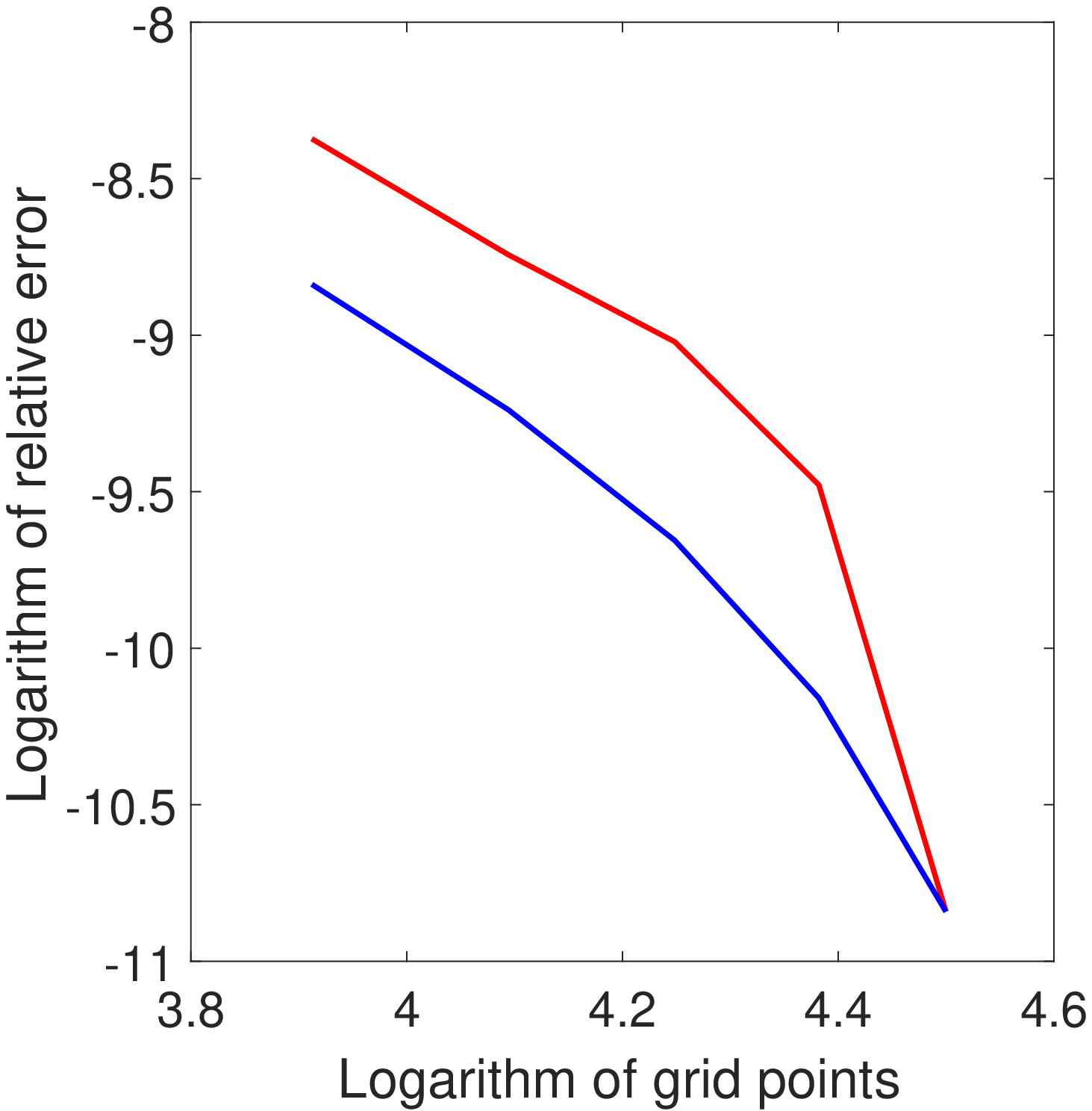}
}
 \subfigure[]{\label{subfig-b-hestonsabr}
 \includegraphics[width=4.5cm]{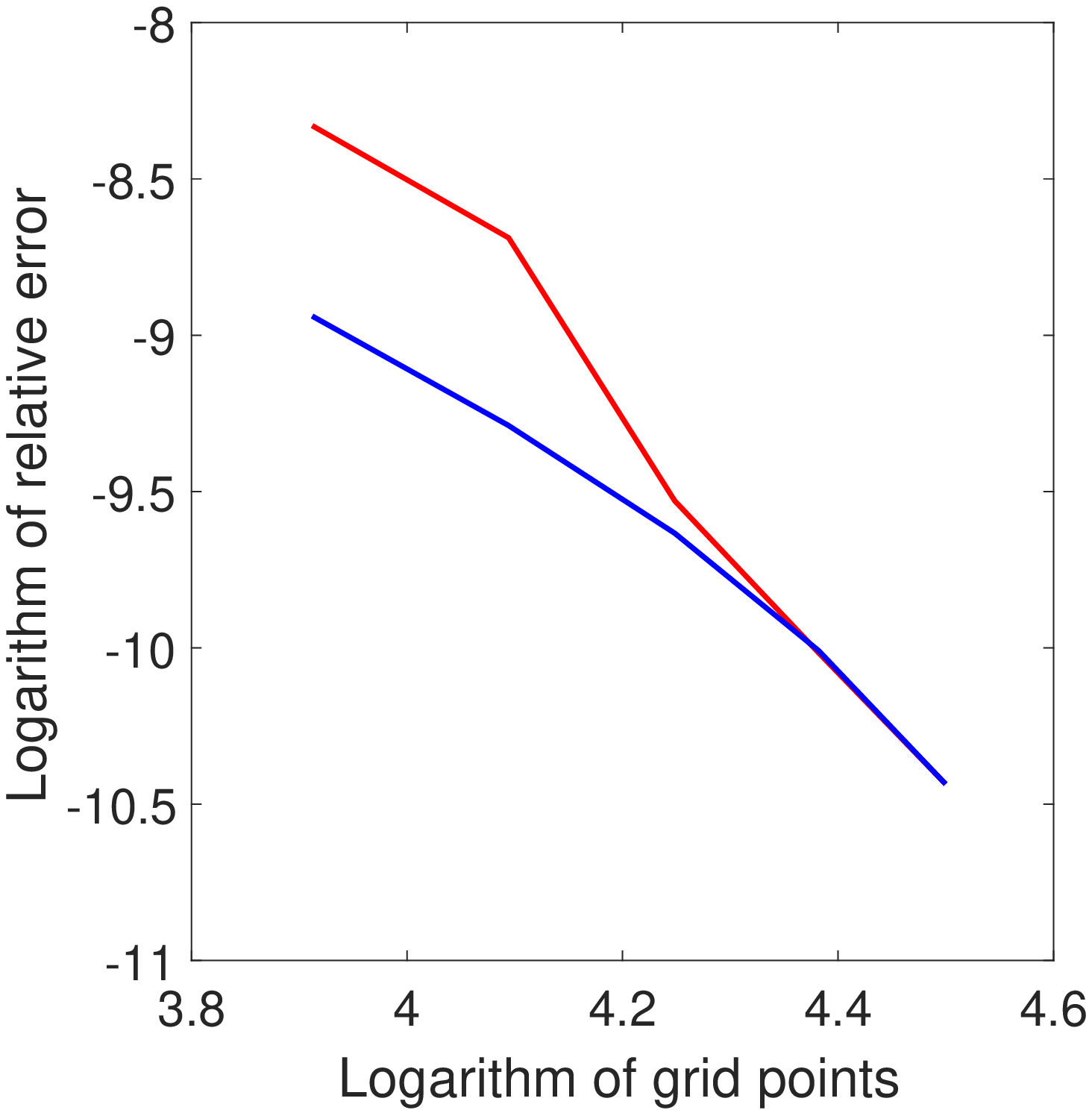}
}
 \subfigure[]{\label{subfig-b-qslv}
 \includegraphics[width=4.5cm]{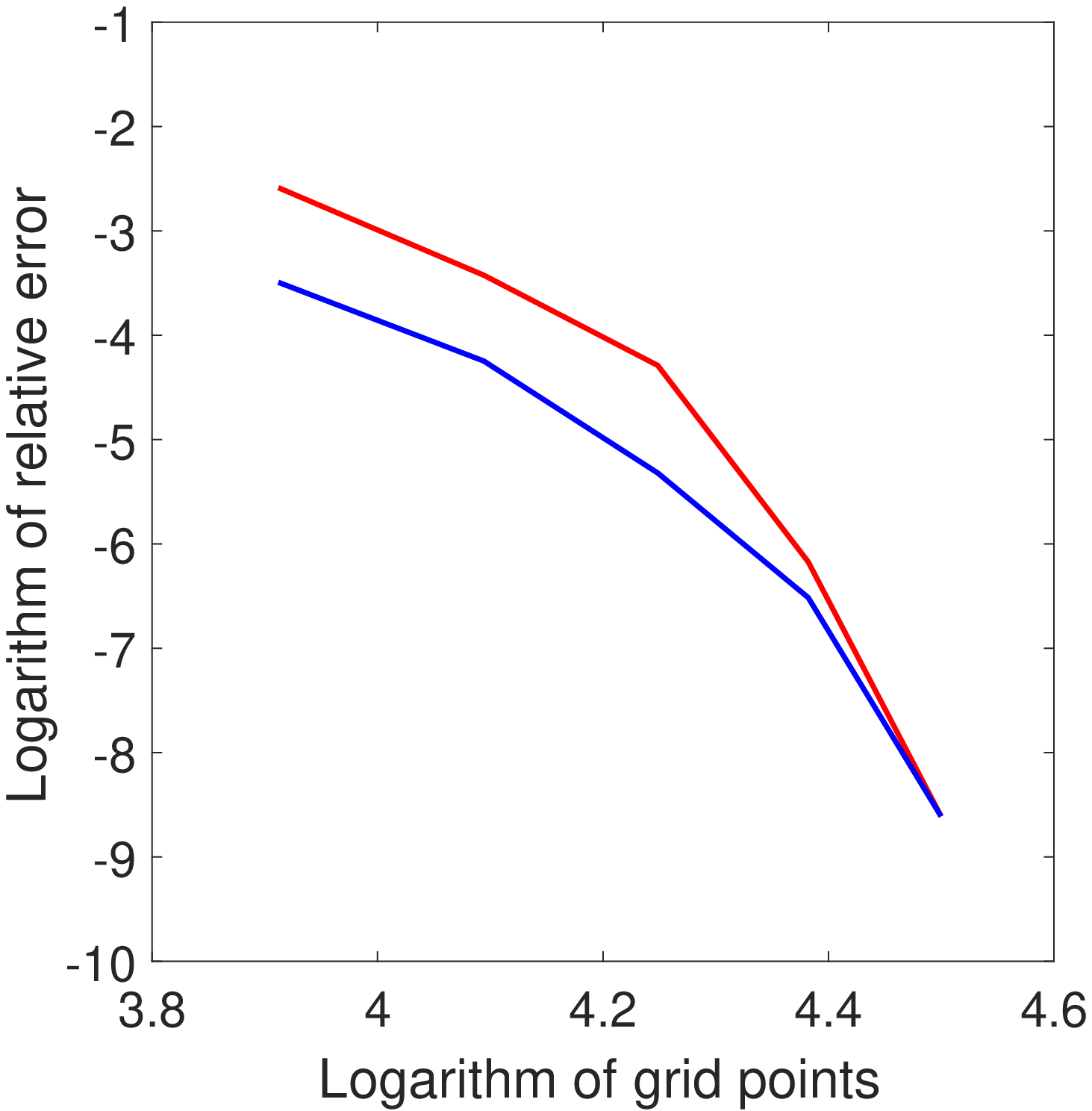}
}
 \caption{The logarithmic plot of the number of grid points versus relative error for barrier options under rough stochastic local volatility models via Algorithm \ref{alg:fast}.  Figures \ref{subfig-b-heston}, \ref{subfig-b-fourtwo}, \ref{subfig-b-alpha}, \ref{subfig-b-sabr}, \ref{subfig-b-hestonsabr}, \ref{subfig-b-qslv} correspond to rough Heston, $4/2$, $\alpha$-Hyper, SABR, Heston-SABR, quadratic SLV  models, respectively. The red lines are obtained by fixing $M=90$, and then increasing the grid points of $N$ from $50$ to $90$. The blue ones are obtained by fixing $N=90$ and changing $M$. Here we take the prices obtained by $M=N=100$, $\varepsilon=10^{-10}$ as the benchmarks to calculate relative errors. }\label{fig-barrier}
\end{figure}

Table \ref{tab-european-epsilon} shows that with the decrease of $\varepsilon$, the European options prices under RSLV models will converge to the benchmarks. For barrier options, there are similar results, which are listed in Table \ref{tab-barrier-epsilon}. These results show the accuracy of the semimartingale and CTMC approximation algorithm. From Figures \ref{fig-European} and  \ref{fig-barrier}, we see that for a fixed $\varepsilon$, increasing the number of CTMC grids will make the relative error between the calculated results and the benchmark gradually decrease.  It is worth mentioning that, for fixed $N=M=100$, the average CPU times to calculate the European and barrier option prices are respectively $0.18$ and $0.20$ seconds.  This shows the very high efficiency of our algorithm.
\end{example}

\begin{table}[htbp]
\centering
\caption{American call option under the RSLV models.}\label{tab-american-epsilon}
\begin{threeparttable}
\begin{tabular}{l l l l l l l}
 \toprule
 \multirow{2}*{$\varepsilon$} & \multicolumn{5}{c}{Relative Errors}\\
\cline{2-7}
&R-H &R-$4/2$ &R-$\alpha$-H &R-SABR &R-H-SABR &R-Q-SLV\\
\hline
$10^{-4}$  &9.82e-3  &9.99e-3  &8.53e-3 &9.25e-3 &7.41e-3  & 8.88e-3\\
\hline
$10^{-5}$  &6.73e-3  &9.36e-3  &6.25e-3 &8.66e-3 &5.55e-3  & 7.21e-3\\
\hline
$10^{-6}$  &4.21e-3  &8.95e-3  &4.33e-3 &7.99e-3 &4.21e-3  & 6.85e-3\\
\hline
$10^{-7}$  &3.75e-3  &8.50e-3  &3.28e-3 &7.21e-3 &3.48e-3  & 6.01e-3\\
\hline
$10^{-8}$  &2.88e-3  &8.12e-3  &2.79e-3 &6.54e-3 &3.01e-3  & 5.75e-3\\
\hline
benchmark &6.0635  &0.0418 &6.1111 &6.0000 &6.4410 & 7.1658\\
 \bottomrule
\end{tabular}
\begin{tablenotes}
\item[*] Here ``R-H", ``R-$4/2$", ``R-$\alpha$-H", ``R-SABR", ``R-H-SABR", ``R-Q-SLV" represent ``rough Heston", ``rough $4/2$ model", ``rough $\alpha$-Hyper", ``rough SABR", ``rough Heston-SABR", ``rough rough quadratic SLV" models respectively. Benchmarks obtained by fast simulation algorithm based on least squares Monte Carlo method with $10^5$ simple paths, and they take about $875.26$ seconds on average. The results in the table are calculated with $N=M=100$ via Algorithm \ref{alg:American}, and they take $91.24$ seconds on average.
    \end{tablenotes}
    \end{threeparttable}
\end{table}

\begin{figure}[htbp]
\centering
 \subfigure[]{\label{subfig-A-heston}
 \includegraphics[width=4.5cm]{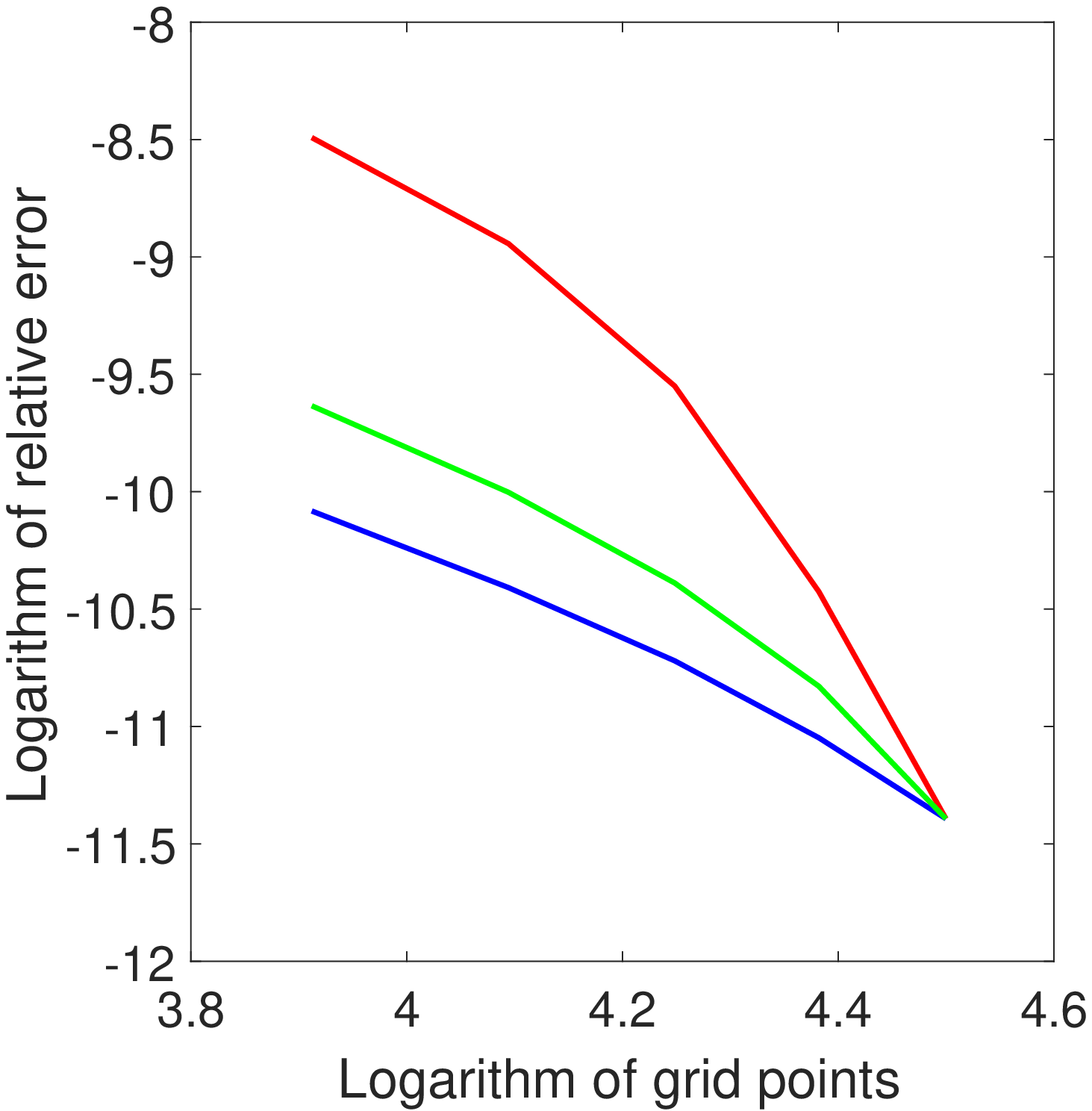}
}
 \subfigure[]{\label{subfig-A-fourtwo}
 \includegraphics[width=4.5cm]{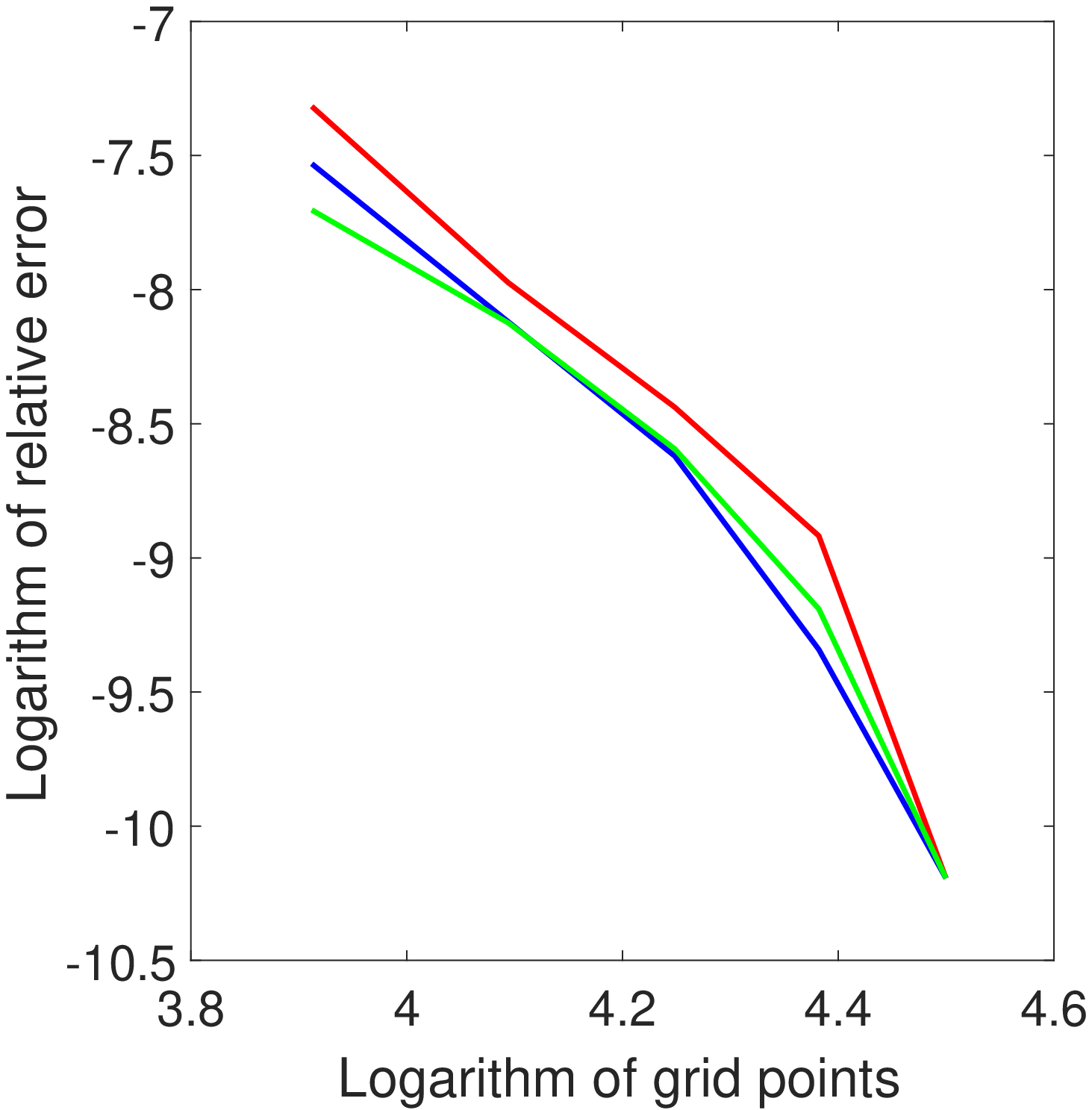}
}
 \subfigure[]{\label{subfig-A-alpha}
 \includegraphics[width=4.5cm]{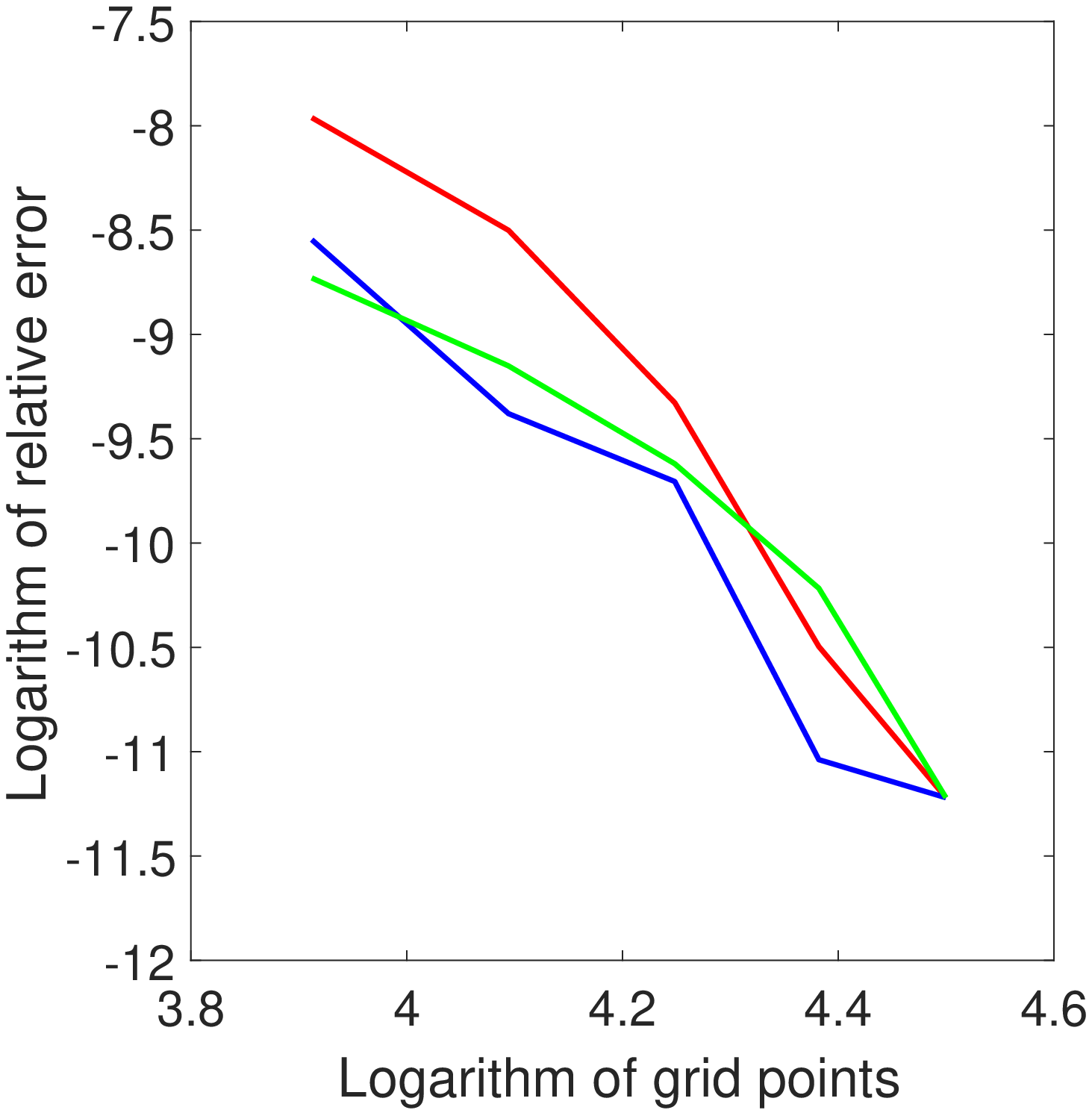}
}
 \subfigure[]{\label{subfig-A-sabr}
 \includegraphics[width=4.5cm]{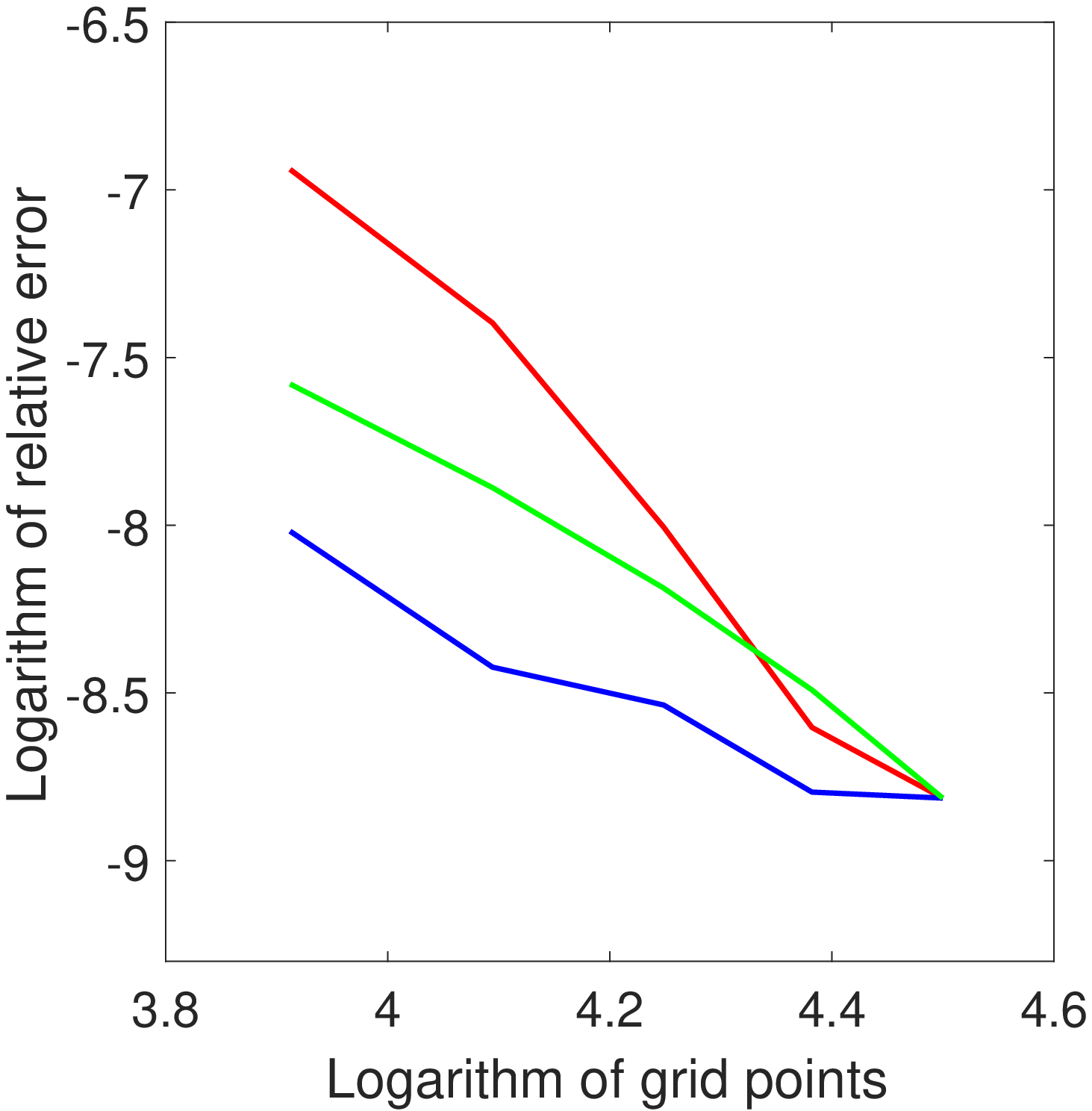}
}
 \subfigure[]{\label{subfig-A-hestonsabr}
 \includegraphics[width=4.5cm]{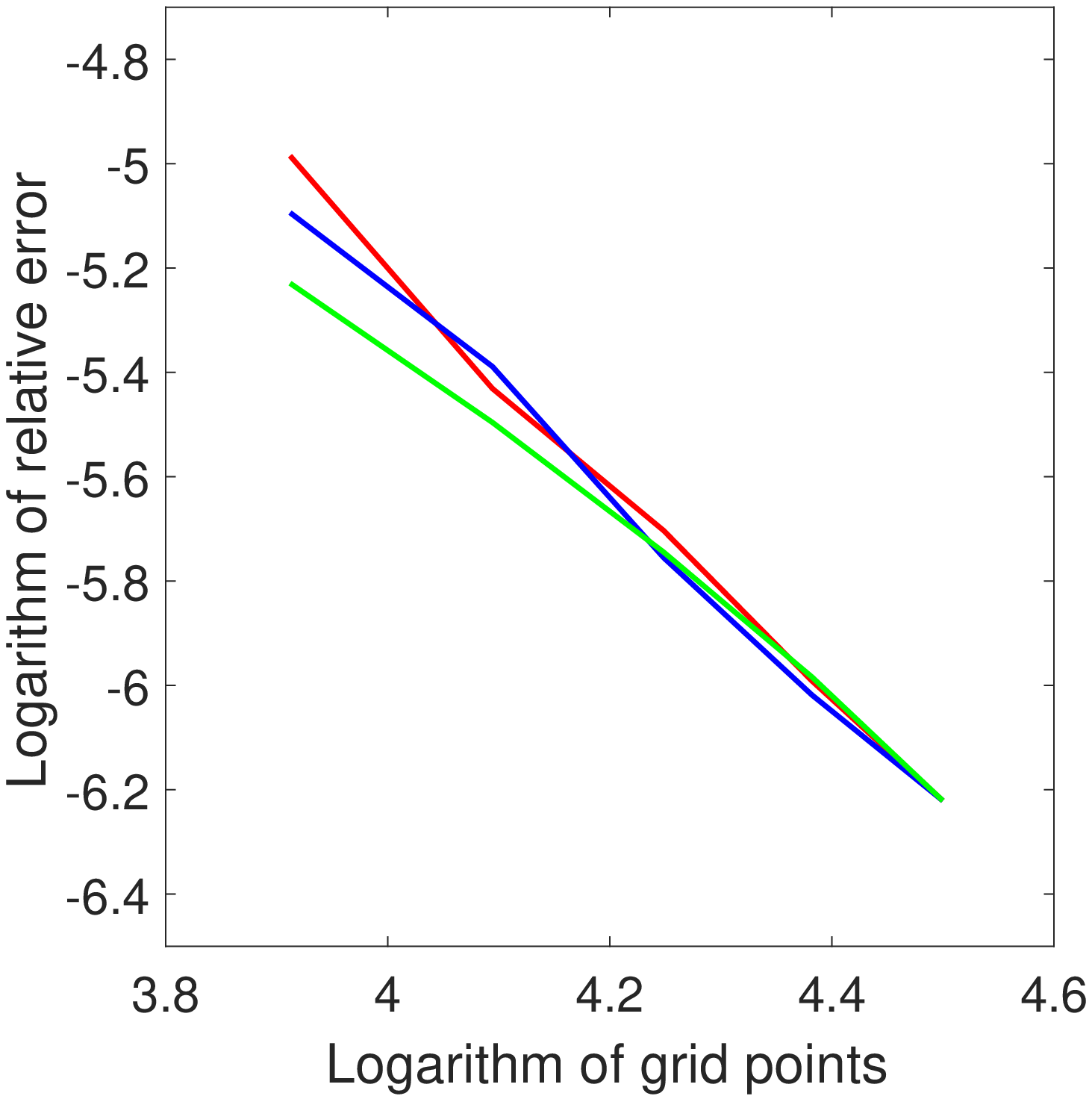}
}
 \subfigure[]{\label{subfig-A-qslv}
 \includegraphics[width=4.5cm]{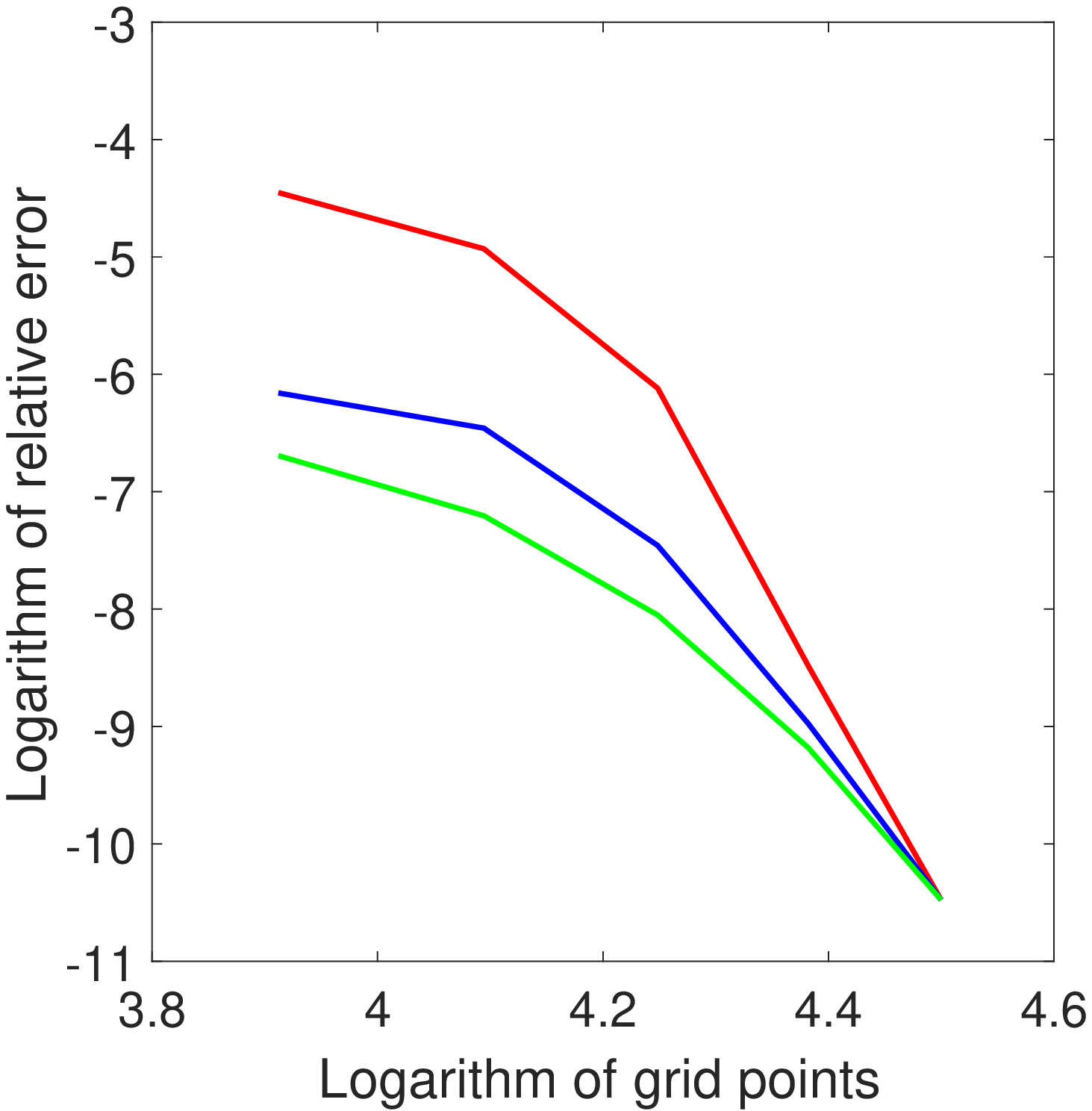}
}
 \caption{The logarithmic plot of the number of grid points versus relative error for American options under rough stochastic local volatility models via Algorithm \ref{alg:American}.  Figures \ref{subfig-A-heston}, \ref{subfig-A-fourtwo}, \ref{subfig-A-alpha}, \ref{subfig-A-sabr}, \ref{subfig-A-hestonsabr}, \ref{subfig-A-qslv} correspond to rough Heston, $4/2$, $\alpha$-Hyper, SABR, Heston-SABR, quadratic SLV  models, respectively. The red lines are obtained by fixing $n=M=90$, and then increasing the grid points of $N$ from $50$ to $90$. The blue ones are obtained by fixing $n=N=90$ and changing $M$, and the green ones are obtained by fixing $N=M=90$ and changing $n$. Here we take the prices obtained by $n=N=M=100$, $\varepsilon=10^{-10}$ as the benchmarks to calculate relative errors. }\label{fig-American}
\end{figure}

\newpage

\begin{example}[American option]
In this example, we use Algorithm \ref{alg:American} to calculate the price of American options under the six models listed in the Table \ref{tab-rslv-model}.

Table \ref{tab-american-epsilon} shows that with the decrease of $\varepsilon$, the American options prices under RSLV models will converge to the benchmark. This demonstrates the accuracy of the semimartingale and CTMC approximation algorithm. In addition to showing the convergence results of CTMC grid points $N$ and $M$, Figure \ref{fig-American} also shows the influence of the change in grid points on the relative error in the time direction. These results show the convergence of Algorithm \ref{alg:American}. Since the value of the American option depends on the optimal stopping time, the fast Algorithm \ref{alg:fast} is not available. However, the average CPU time to calculate American option prices is $91.24$ seconds, which is significantly less than the  least squares Monte Carlo method. Moreover, compared with Table 4 in \citet{goudenege}, which spend a significant amount of time to price American options under the simple Bergomi model,  Algorithm \ref{alg:American} is clearly more effective and widely applicable for pricing American options under general rough stochastic local volatility models.
\end{example}

\section{Conclusions}\label{section-conclusion}
In this paper, we first propose a perturbation approach to approximate the rough stochastic local volatility model by a perturbed stochastic local volatility model, which is a semimartingale. After further expressing it as a Markovian form, we propose a continuous-time Markov chain approximation approach to derive  the semi-explicit expressions of European, barrier and American options prices. The approximate expression obtained by this method is proved to converge to the solution of the original  option pricing problem under the RSLV model. Numerical experiments demonstrate the accuracy and very high efficiency of the method for several RSLV models, including rough Heston and rough SABR models, etc.


\end{document}